\documentclass[final,5p,times,twocolumn]{elsarticle}
\usepackage[breaklinks]{hyperref}
\usepackage{xcolor}
\usepackage{tikz}
\usepackage{listings}
\usepackage{graphicx}
\graphicspath{{./fig/}}
\usepackage{etoolbox} % fix citation spacing issue
\apptocmd{\thebibliography}{\raggedright}{}{}
\usepackage{caption}
\usepackage{float}
\usepackage{microtype}
\usepackage{subcaption}
\usepackage{booktabs}
\usepackage{tikz}
\usetikzlibrary{shapes.geometric, arrows, decorations.pathreplacing, angles, quotes, arrows.meta, calc, shadings}
\usepackage{amssymb}
\usepackage{amsmath}
\journal{Computer Physics Communications}

\begin{document}

\begin{frontmatter}



  \title{Geant4 Optical Simulation without C++}


  \author{Ariestotle Raj Maharjan}
  \author{Jianchen Li}
  \author{Jing Liu}
  \ead{jing.liu@usd.edu}
  \affiliation{organization={Department of Physics, University of South Dakota},%Department and Organization
    addressline={414 E. Clark St.},
    city={Vermillion},
    postcode={57069},
    state={SD},
  country={USA}}

  %% Abstract
  \begin{abstract}
    The plain text geometry description syntax in Geant4 has been extended to incorporate optical properties for bulk materials and surface interfaces. This extension enables users to configure and execute comprehensive optical simulations without writing C++ code, significantly lowering the learning curve and eliminating the need for frequent recompilation. In this paper, we detail the implementation of the new \texttt{:prop} and \texttt{:surf} tags and validate them through examples of key optical processes, including Cherenkov radiation, scintillation, Rayleigh scattering, and absorption. Furthermore, we provide a thorough demonstration of configuring complex optical boundaries using the UNIFIED model. These capabilities are contextualized through practical scenarios, showcasing the extension's potential for rapid prototyping and simulation studies.

    \noindent \textbf{Program Summary} \\
    \emph{Program Title:} GEARS - Geant4 Example Application with Rich features and Small footprint\\
    \emph{Developer's repository link:} \url{https://github.com/jintonic/gears} \\
    \emph{Licensing provisions:} MIT \\
    \emph{Programming language:} C++ \\
    \emph{External routines/libraries:} Geant4, CMake \\
    \emph{Nature of problem:} Configuring Geant4 optical simulations traditionally requires advanced C++ programming, creating a steep learning curve and necessitating time-consuming recompilation during the iterative detector design process.\\
    \emph{Solution method:} We extended the Geant4 plain text geometry syntax with custom tags for defining bulk and surface optical properties. These text-based definitions are parsed at runtime, enabling fast, C++-free configuration of complex optical simulations.
  \end{abstract}


\end{frontmatter}

%% Use \subsubsection, \paragraph, \subparagraph commands to
%% start 3rd, 4th and 5th level sections.
%% Refer following link for more details.
%% https://en.wikibooks.org/wiki/LaTeX/Document_Structure#Sectioning_commands

\section{Introduction}

Geant4~\cite{g403, g406, g416} is a widely used toolkit for simulating the passage of particles through matter, including optical photons.  It provides three methods to define detector geometry and material: C++, GDML~\cite{gdml06}, and plain text with some simple syntax~\cite{tg}. All of them are introduced in the \emph{ Detector Definition and Response} section in \emph{Geant4 User's Guide for Application Developers} (Geant4 User's Guide in short hereafter)~\cite{g4doc} with an obvious emphasis on C++.

\subsection{C++ VS GDML \& Plain Text}

However, detector definition in C++ requires fair amount of knowledge of C++ programming and the usage of Geant4 classes. This creates a steep learning curve for beginning physicists and engineers. In addition, it requires re-compilation every time a change is made, no matter how small. This is very inconvenient, especially during the iterative design and debugging phase.

In contrast, GDML and plain text geometry definitions do not require compilation at all, allowing for fast and frequent updates to a geometry setup.

\subsection{GDML VS Plain Text}

GDML is akin to HTML in that it uses a structured and verbose XML-based syntax, which provides greater flexibility and precision but requires familiarity with its syntax rules. Plain text geometry description, on the other hand, is more like Markdown~\cite{md}: simple, intuitive, and easier to write and read, making it ideal for quick prototyping and debugging.

In fact, GDML is created as an intermediate format to facilitate the exchange of geometry information between different tools, such as FreeCAD~\cite{cad}, Geant4, and ROOT~\cite{root}. It is meant to be generated by one piece of software and read by another. The strict and structured XML-based syntax is well suited for this purpose. However, its complexity can be a barrier for users who want to quickly define and modify geometries without delving into the intricacies of XML syntax. On the other hand, plain text geometry description is designed to be written by human, allowing users to easily create and modify geometries without the need for extensive knowledge of XML or programming.

\subsection{Optical Simulation}

Optical simulation in Geant4 typically requires C++ coding to define optical properties of a bulk material and/or an interface (surface) between two materials~\cite{g4doc}. However, it suffers the same drawbacks described previously. Optical properties can also be defined using GDML with, however, some limitation, for example, optical properties cannot be assigned to existing NIST materials pre-defined in Geant4~\cite{gdmllimit}, such as \verb|G4_AIR|, etc.

Although the standard plain text geometry definition in Geant4 lacks syntax for optical properties, we have extended it to include them for both bulk materials and surfaces. This allows users to perform optical simulations without C++ coding, making the process more accessible and user-friendly.

\subsection{Article Structure}

In this article, we first introduce our extension to the plain text geometry definition syntax for defining optical properties in Section~\ref{s:tg}, and then demonstrate its usage and accuracy in simulating optical properties of bulk materials (Section~\ref{s:mat}) and optical interfaces (Section~\ref{s:sur}), individually. The demonstration is done primarily through the application of this extension to scintillation crystals wrapped with various reflectors and read out by common photo-sensors, such as photomultiplier tubes (PMTs) or silicon photomultipliers (SiPMs). Finally, we summarize the work and discuss its implications in the concluding section.

\section{Plain Text Extension}\label{s:tg}

The plain text geometry definition in Geant4 utilizes tags (keywords with a proceding colon) to define various aspects of the geometry, such as \verb|:solid| for a 3D shape, \verb|:mate| for a material, and \verb|:volu| for a Geant4 logical volume. For example,
\begin{lstlisting}[basicstyle=\footnotesize]
:volu CsI TUBE 0 3*cm 5*cm G4_CESIUM_IODIDE
\end{lstlisting}
defines a logical volume named \verb|CsI| as a tube with an inner radius of 0, an outer radius of 3 cm, a half-length of 5 cm, and made of the built-in NIST material \verb|G4_CESIUM_IODIDE|.

We have added two new tags to define optical properties for bulk materials (\verb|:prop|) and surfaces (\verb|:surf|), following the instruction in section \emph{Defining new tags in the geometry file format} in the plain text geometry definition manual~\cite{tg}. The C++ code is included in a single-file Geant4 application, GEARS - Geant4 Example Application with Rich features and Small foodprint~\cite{gears}, that can parse the extended plain text geometry description in an ASCII file (\verb|detector.tg|, for example), and set up the optical simulation accordingly. A new Geant4 macro command, \verb|/geometry/source|, is also introduced in GEARS to specify the geometry definition file:
\begin{lstlisting}[emph={geometry, source}, emphstyle=\bfseries]
/geometry/source detector.tg
\end{lstlisting}

\subsection{New Tag :prop}
The \verb|:property| (or \verb|:prop| in short) tag is introduced to define both constant and energy dependent optical properties of a bulk material:
\begin{lstlisting}
:prop [material]
 [property_a] [parameter]
 photon_energies [size] [array]
 [property_b] [parameter array]
\end{lstlisting}
For example:
\begin{lstlisting}[caption=Optical properties defined in file CsI77K.tg., label=l:csi]
:prop G4_CESIUM_IODIDE
 ScintillationYield 100/keV
 ResolutionScale 4.5
 ScintillationYield1 0.02
 ScintillationYield2 0.98
 ScintillationTimeConstant1 2*ns
 ScintillationTimeConstant2 1*us
 photon_energies 101 2.7*eV ... 4.8*eV
 ScintillationComponent1 0 ... 1.487e-05
 ScintillationComponent2 1.56e-05 ... 0
 Rindex 1.82 ... 2.146
 AbsLength 30*cm ... 30*cm
 Rayleigh 339*cm ... 339*cm
\end{lstlisting}

Note that the material specified in the \verb|:prop| tag must be defined previously using the \verb|:mate| tag or be a built-in NIST material in Geant4, such as \verb|G4_CESIUM_IODIDE|.

Both tags and properties are case-insensitive. Properties that contains more than one word, such as \verb|ScintillationYield|, can be written in camel case~\cite{camel} to increase readability.

A list of optical properties and their meanings can be found in section \emph{Optical Photon Processes} in Geant4 User's Guide~\cite{g4doc}.

\subsection{New Tag :surf}
The \verb|:surface| (or \verb|:surf| in short) tag is introduced to define optical properties for a surface (interface between two physical volumes):

\begin{lstlisting}
:surf name [vol.1]:[#] [vol.2]:[#]
  type [...]
  model [...]
  finish[...]
  sigma_alpha [...]
    property
    [property_a] [value]
    photon_energies [size] [array]
    [property_b] [parameter array]
\end{lstlisting}
where \verb|[vol.1]:[#]| means the name of physical volume 1 and its copy number \#~\cite{g4doc}. Copy number is necessary to uniquely identify a specific placement of a volume that may be placed multiple times. For example:
\begin{lstlisting}[caption={Definition of an optical surface using the UNIFIED model.}, label=l:surf]
:surf CsI_Teflon CsI:1 Teflon:2
  model unified
  type dielectric_dielectric
  finish groundBackPainted
  sigma_alpha 0.1
    property
    photon_energies 2 2.5*eV 5.0*eV
    Reflectivity 0.9 0.8
    Rindex 1.0 1.1
\end{lstlisting}
Note that properties specified in the \verb|:surf| tag are assigned to the interface between two physical volumes identified by their names and copy numbers, and they must be defined at the end of the \verb|:surf| code block after the \emph{property} keyword.

\subsection{Usage}
To use the extension, a user needs to provide at least two text files to GEARS: 1. a detector definition file, for example, \verb|detector.tg|, with optical properties defined using the new tags, and 2. a macro file, for example \verb|run.mac|, generating optical photons in the detector system.

Listing~\ref{l:run} shows a minimal working example of \verb|run.mac|. It instructs Geant4 to simulate 1000 times a 3.5~eV optical photons emitted isotropically from the center of a detector defined in \verb|detector.tg|.

\begin{lstlisting}[emph={geometry, source, physics_lists, run, gps, energy, particle, factory, initialize, beamOn, ang, type, addOptical, polarization},
emphstyle=\bfseries, caption=Minimal working example of Geant4 macro file run.mac., label=l:run]
/geometry/source detector.tg
/physics_lists/factory/addOptical
/run/initialize
/gps/particle opticalphoton
/gps/polarization 1 0 0
/gps/energy 3.5 eV
/gps/ang/type iso
/run/beamOn 1000
\end{lstlisting}

It is possible to pull optical property definitions out of \verb|detector.tg| and place them in a separate file, for example, \verb|CsI77K.tg|, as shown in Listing~\ref{l:csi}, which can be included back in \verb|detector.tg| as shown in Listing~\ref{l:det}.

\begin{lstlisting}[caption=Example detector.tg file including CsI77K.tg, label=l:det]
#include /path/to/CsI77K.tg
:volu CsI TUBE 0 30 50 G4_CESIUM_IODIDE
\end{lstlisting}

The advantage of this approach is that the optical property definition file \verb|CsI77K.tg| can be reused in various detector definition files.

When parsing the optical definition code, GEARS will print out the following messages on screen for the user to verify whether the definitions are correctly recognized.

\begin{lstlisting}[basicstyle=\small]
GEARS: !Set optical properties of! G4_CESIUM_IODIDE:
GEARS: SCINTILLATIONYIELD=100/keV
GEARS: RESOLUTIONSCALE=4.5
GEARS: SCINTILLATIONTIMECONSTANT1=2*ns
GEARS: SCINTILLATIONTIMECONSTANT2=1*us
GEARS: SCINTILLATIONYIELD1=0.02
GEARS: SCINTILLATIONYIELD2=0.98
GEARS: SCINTILLATIONCOMPONENT1=0, 0...
GEARS: SCINTILLATIONCOMPONENT2=1.556e-05, ...
GEARS: RINDEX=1.82, 1.822...
GEARS: ABSLENGTH=300, 300...
GEARS: RAYLEIGH=3390, 3390...
\end{lstlisting}

In the following sections, we demonstrate the application of the extended plain text geometry definition through various example snippets of geometry definition files, macro files, and simulation results.

\section{Material Optical Properties} \label{s:mat}

The effectiveness of configuring optical properties of a bulk material using the \verb|:prop| tag is verified through a few key optical properties defined in the \emph{Optical Photon Processes} section~\cite{g4doc} in Geant4 User's Guide. Some of them are for the generation of optical photons, such as scintillation and Cherenkov radiation, while others are for the transportation of optical photons, such as absorption length and Rayleigh scattering length.

\subsection{Cherenkov Radiation}

\begin{figure}[htbp]\centering
  \includegraphics[width=0.9\linewidth]{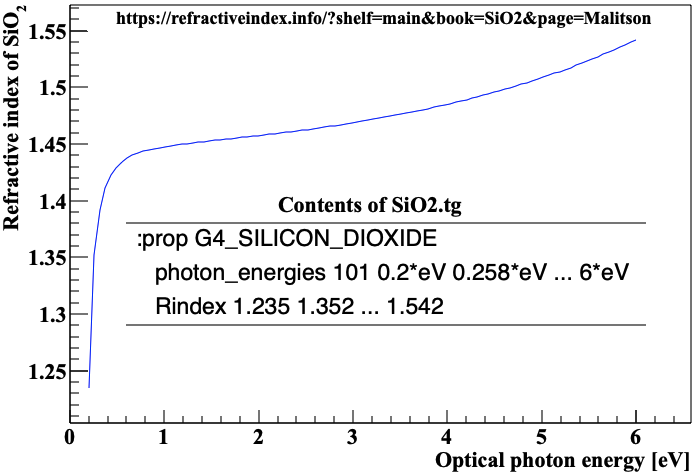}
  \caption{Refractive index of silicon dioxide (SiO$_2$) defined in SiO2.tg file.}
  \label{f:nwin}
\end{figure}

Cherenkov radiation is emitted when a charged particle travels through a dielectric medium at a speed greater than the phase velocity of light in that medium. The refractive index of the medium is a key parameter that determines the Cherenkov threshold and the angle of emission~\cite{g4phy}. In general, the refractive index is energy dependent, and it is defined using the \emph{Rindex} property in the \verb|:prop| tag. Figure~\ref{f:nwin} shows the refractive index of silicon dioxide (SiO$_2$) taken from Ref.~\cite{n}. Saved in file \verb|SiO2.tg| are 101 data points along this curve as shown in the code snippet in Figure~\ref{f:nwin}. This file can then be included in a detector definition file, for example, \verb|PMTwindow.tg|, to define the refractive index of SiO$_2$, a common material used in PMT windows, as shown in Listing~\ref{l:pmt}.

\begin{lstlisting}[caption={Snippet of PMTwindow.tg file that uses SiO2.tg.}, label=l:pmt]
#include /path/to/SiO2.tg
:volu window TUBE 0 38.1 1 G4_SILICON_DIOXIDE
\end{lstlisting}

\verb|PMTwindow.tg| can then be used by a Geant4 macro file, for example, \verb|cherenkov.mac|, to simulate Cherenkov radiation emitted by a 511~keV electron passing through a PMT window, as shown in Listing~\ref{l:cherenkov}.

\begin{lstlisting}[caption=cherenkov.mac to simulate Cherenkov radiation in a PMT window., label=l:cherenkov, numbers=left, numbersep=5px, numberstyle=\color{blue}, basicstyle=\footnotesize, emph={geometry, source, physics_lists, run, gps, energy, particle, factory, initialize, beamOn, addOptical, tracking, verbose, position}, emphstyle=\bfseries]
/geometry/source /path/to/PMTwindow.tg
/physics_lists/factory/addOptical
#/process/optical/processActivation Cerenkov false
/run/initialize
/gps/particle e-
/gps/energy 511 keV
/gps/position 0 0 1.01 mm
/tracking/verbose 2
/run/beamOn
\end{lstlisting}

\begin{figure*}[t]
\begin{lstlisting}[caption=Cherenkov photons emitted as secondary particles through an electron multiple scattering (msc) process., label=l:crkv, frame=none, basicstyle=\footnotesize]
*********************************************************************************************************
* G4Track Information:   Particle = e-,   Track ID = 1,   Parent ID = 0
*********************************************************************************************************
Step#    X(mm)    Y(mm)    Z(mm) KinE(MeV)  dE(MeV) StepLeng TrackLeng  NextVolume ProcName
    0        0        0     1.01       0.5        0        0         0      vacuum initStep
    1        0        0        1       0.5 3.05e-28     0.01      0.01      window CoupledTransportation
    2  0.00152  0.00458     0.97     0.492  0.00782   0.0309    0.0409      window msc
    :----- List of 2ndaries - #SpawnInStep=  3(Rest= 0,Along= 0,Post= 3), #SpawnTotal=  3 ---------------
    :  0.000866   0.00261     0.983  3.32e-06      opticalphoton
    :  0.000704   0.00212     0.986  6.96e-07      opticalphoton
    :  7.96e-05   0.00024     0.998   4.1e-06      opticalphoton
    :----------------------------------------------------------------- EndOf2ndaries Info ---------------
\end{lstlisting}

\begin{lstlisting}[caption=No Cherekov photon emission from msc when Chereknov process is disabled., label=l:msc, frame=none, basicstyle=\footnotesize]
*********************************************************************************************************
* G4Track Information:   Particle = e-,   Track ID = 1,   Parent ID = 0
*********************************************************************************************************
Step#    X(mm)    Y(mm)    Z(mm) KinE(MeV)  dE(MeV) StepLeng TrackLeng  NextVolume ProcName
    0        0        0     1.01       0.5        0        0         0      vacuum initStep
    1        0        0        1       0.5 3.05e-28     0.01      0.01      window CoupledTransportation
    2  -0.0064 0.000274    0.964      0.48   0.0202   0.0374    0.0474      window msc
    ...
    7 -0.00554   -0.165    0.597     0.267    0.122    0.198     0.493      window msc
    8    -0.03    -0.12    0.459         0    0.267    0.368     0.861      window eIoni
    9    -0.03    -0.12    0.459         0        0        0     0.861      window NoProcess
\end{lstlisting}
\end{figure*}

\begin{figure}[htbp]\centering
  \includegraphics[width=0.9\linewidth]{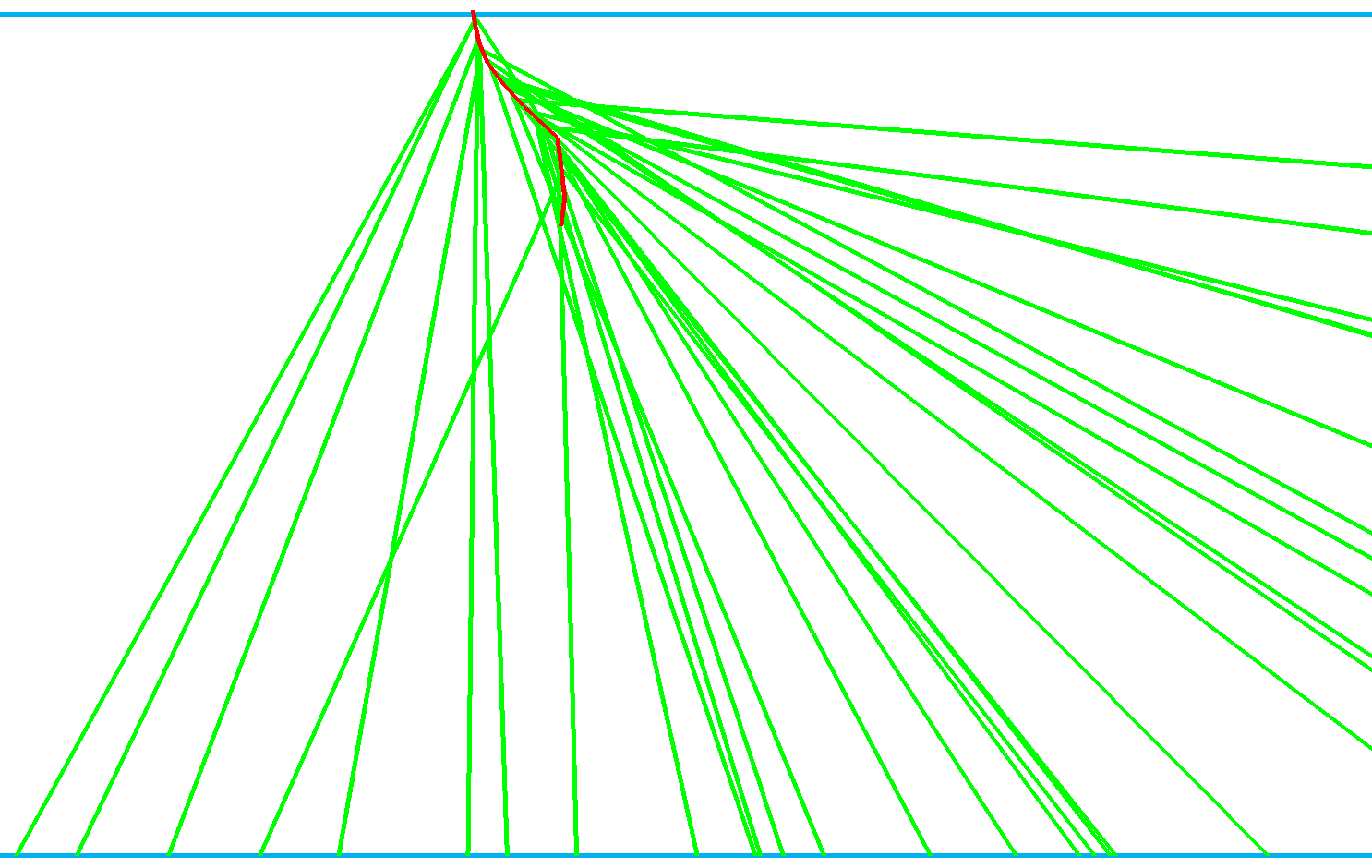}
  \caption{Cherenkov photons (green) emitted along an electron trajectory (red) in a PMT window (blue).}
  \label{f:crkv}
\end{figure}

The simulation can be performed by executing command \verb|gears| \verb|cherenkov.mac| in a terminal. Figure~\ref{f:crkv} shows the visualization of Cherenkov photons (green) emitted along an electron trajectory (red) in a PMT window (blue). The detailed step-by-step tracking information can be printed out on screen for verification using Geant4 macro command \verb|/tracking/verbose 2|, as shown in Listing~\ref{l:crkv}.

Note that optical photons are generated as secondary particles through an electron multiple scattering (msc) process instead of an explicit Cherenkov process. To verify that they are really Cherenkov photons, one can disable the Cherenkov radiation process by uncommenting line 3 in Listing~\ref{l:cherenkov}, and re-run the simulation. The output tracking information, as shown in Listing~\ref{l:msc}, confirms that no optical photons can be generated during the electron multiple scattering process if the Cherenkov process is disabled.

\begin{figure}[htbp]
  \centering
  \includegraphics[width=\linewidth]{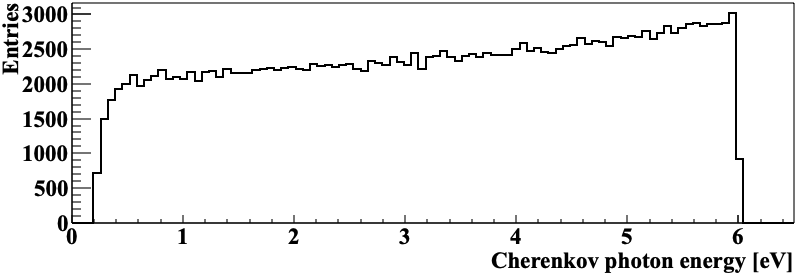}
  \caption{Cherenkov photon energy spectrum in a PMT window.}
  \label{f:crkve}
\end{figure}

Figure~\ref{f:crkve} shows the energy spectrum of Cherenkov photons generated in the PMT window. The artificial cut-off of the spetrum at 6~eV is due to the lack of refractive index data above 6~eV in \verb|SiO2.tg|. The rest of the spectrum follows roughly the refractive index curve shown in Figure~\ref{f:nwin}. This is because the energy of a photon is sampled from the density function below~\cite{g4phy}:
\begin{equation}
  f(E) \propto 1 - \frac{1}{n^2(E)\beta^2},
\end{equation}
where $E$ is the photon energy, $n(E)$ is the refractive index of the medium at energy $E$, and $\beta$ is the velocity of the charged particle relative to the speed of light in vacuum. The smaller the refractive index, the smaller the density function, and thus fewer photons are generated at that energy.

\subsection{Scintillation}
Geant4 provides many properties to fine-tune the emission of scintillation photons in a bulk material, such as light yield, resolution scale, time constants, and emission spectra, etc.~\cite{g4doc}. As an example, Listing~\ref{l:csi} shows the definition of those properties for a pure CsI crystal at liquid nitrogen temperature (77~K), and Listing~\ref{l:det} shows how this file is used in a detector definition file,  \verb|detector.tg|. Listing~\ref{l:scin} shows snippet of Geant4 macro file \verb|scintillate.mac| used to simulate the emission of scintillation photons generated at the center of the crystal after it absorbs a 1~keV $\gamma$ ray.

\begin{lstlisting}[caption=Code snippet of Geant4 macro file scintillate.mac., label=l:scin, emph={geometry, source, physics_lists, run, gps, energy, particle, factory, initialize, beamOn, addOptical, ang, type}, emphstyle=\bfseries]
/geometry/source detector.tg
/physics_lists/factory/addOptical
/run/initialize
/gps/particle gamma
/gps/energy 1 keV
/gps/ang/type iso
/run/beamOn
\end{lstlisting}

\begin{figure}[h]\centering
  \includegraphics[width=0.45\linewidth]{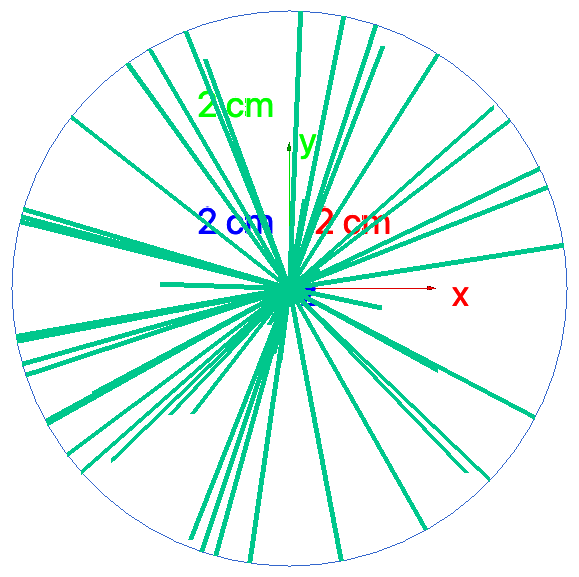}
  \caption{Scintillation photons (green) emitted in a CsI crystal (blue circle).}
  \label{f:sci}
\end{figure}

The simulation can be performed by executing command \verb|gears| \verb|scintillate.mac| in a terminal. Figure~\ref{f:sci} shows the visualization of scintillation photons (green) emitted from the center of a cylindrical CsI crystal (blue circle) after it absorbs a 1~keV $\gamma$ ray, the trajectory of which is too short to be seen. Photon tracks end at the boundary of the crystal because no refractive index is defined beyond the crystal in this simulation. Since Geant4 cannot predict the behavior of optical photons without knowing the refractive index beyond the boundary, it kills the photons upon their arrival at the boundary. Viewing along the axial direction of the cylindrical crystral, some photon tracks appear to be shorter than others because they are killed at the front and back surfaces of the crystal instead of being absorbed in the crystal.

\begin{figure}[htbp]\centering
  \includegraphics[width=0.9\linewidth]{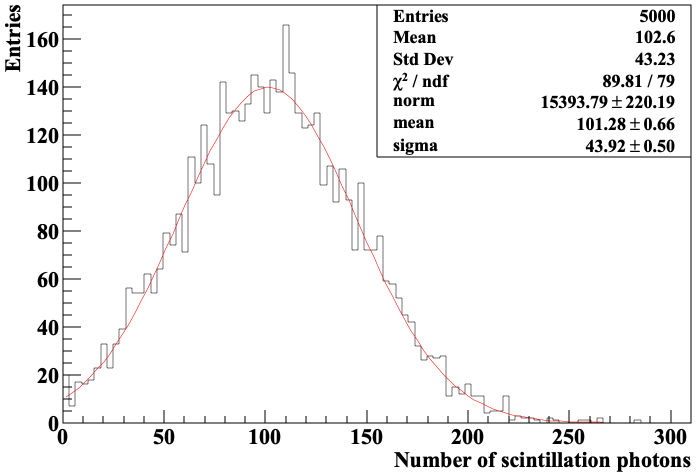}
  \caption{Number of scintillation photons emitted.}
  \label{f:nph}
\end{figure}

Figure~\ref{f:nph} shows numbers of scintillation photons emitted in 5,000 simulated events. It peaks around 101, matching the expected value of 1 keV $\times$ 100/keV within the statistic uncertainty. Assuming Poisson statistics, the width of this distribution should be around $\sqrt{100}=10$. The sigma of the red Gaussian fit is about 4.4 times larger, matching the setting of \verb|ResolutionScale| within the statistic uncertainty.

\begin{figure}[htbp]\centering
  \includegraphics[width=0.9\linewidth]{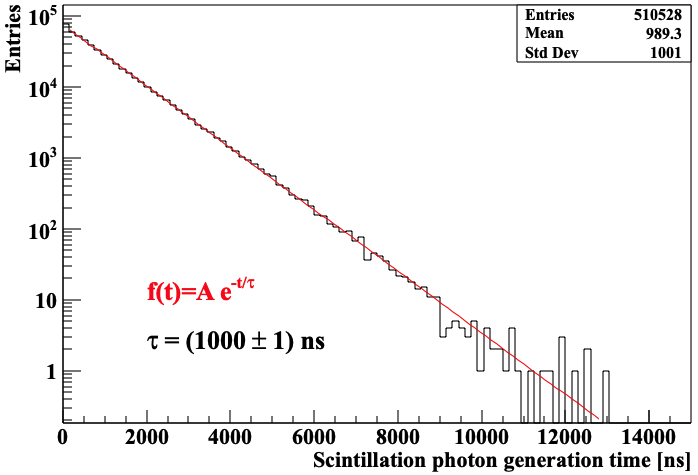}
  \caption{Time distribution of slow scintillation component in CsI crystal.}
  \label{f:tslow}
\end{figure}

\begin{figure}[htbp]\centering
  \includegraphics[width=0.9\linewidth]{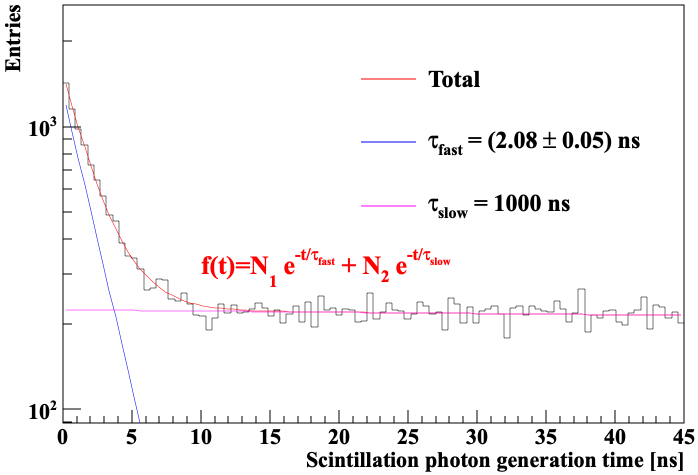}
  \caption{Time distribution of fast scintillation component in CsI crystal.}
  \label{f:tfast}
\end{figure}

Figure~\ref{f:tslow} and \ref{f:tfast} show the emission time distributions of scintillation photons in large and small ranges, respectively. The two exponential decay components are clearly seen, their values obtained from exponential fittings (colored curves) match the time constants defined in Listing~\ref{l:csi} within statistic uncertainties. Note that the slow time constent is fixed to 1000~ns in the fitting of the fast component to improve the fitting stability.

\begin{figure}[htbp]\centering
  \includegraphics[width=0.9\linewidth]{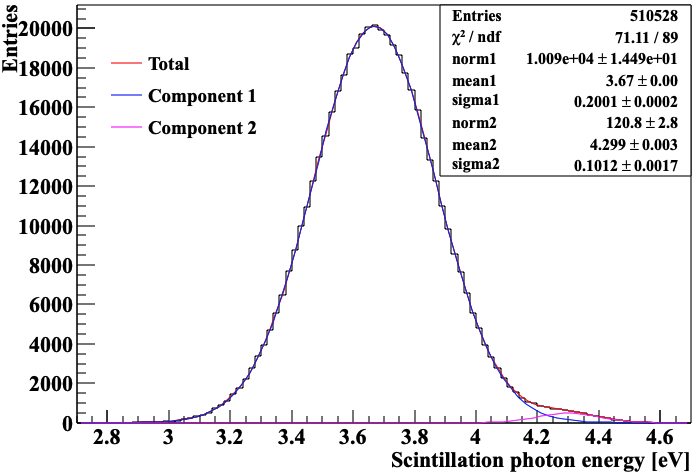}
  \caption{Energy spectrum of scintillation photons in CsI crystal.}
  \label{f:eph}
\end{figure}

Figure~\ref{f:eph} shows the energy spectrum of scintillation photons emitted in the CsI crystal. Two Gaussian functions (blue and magenta) are used to fit the distribution simultaniously. Their means and sigmas match the 3.67 and 4.3~eV emission spectra defined in \verb|CsI77K.tg| (Listing~\ref{l:csi} shows only a snippet of the entire file). The mornalization factors, \emph{norm1} and \emph{norm2}, represent the total areas instead of the heights of the Gaussian distributions. Their ratio is $1.009\times10^4/120.8=84$, much larger than the ratio given by \verb|ScintillationYield2/1| (0.98/0.02=49). This is likely due to a bug in Geant4's implementation of \verb|ScintillationYield1,2|, etc. Further investigation is needed to confirm this. But practically, one can always adjust \verb|ScintillationYield1,2| to achieve the desired ratio of the two emission components.

\subsection{Rayleigh Scattering}
Optical photons can be scattered by small particles in a medium through Rayleigh or Mie scattering depending on the size of particles~\cite{g4doc}. In a typical scintillation crystal, such as CsI, the size of atoms is in the order of a few tenth of nm, much smaller than the typical optical photon wavelength, which is in the order of a few hundred nm. The dominant scattering mechanism hence is Rayleigh scattering.

\begin{figure}[htbp] \centering
  \includegraphics[width=\linewidth]{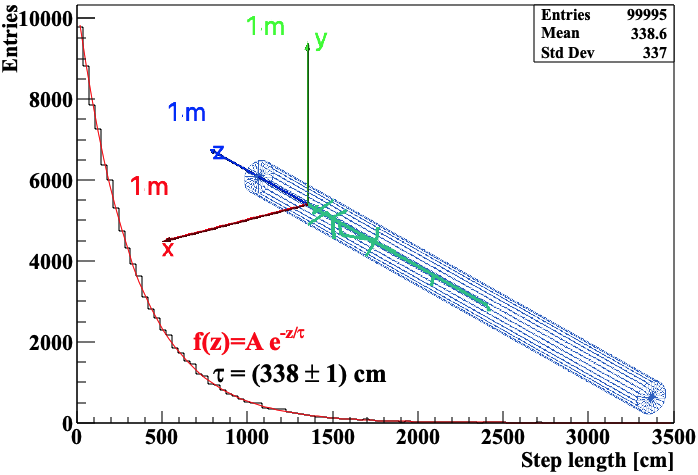}
  \caption{Rayleigh scattering in CsI crystal.}
  \label{f:Rayleigh}
\end{figure}

The Rayleigh scattering length is defined through the property \verb|Rayleigh| in Geant4, as shown in Listing~\ref{l:csi}. To verify the effectiveness of this configuration, 3.5~eV optical photons (green trajectories) are shot along the z axis of a long cylindrical CsI crysatl as shown in the Geant4 visualization imbedded in Figure~\ref{f:Rayleigh}. The distance between the origin and the first scattering point of each trajectory is accumulated in the histogram shown in Figure~\ref{f:Rayleigh}. It is niced described by an exponetial decay function (red) and the decay constant $\tau=338\pm1$~cm given by the fitting matches the set value of 339 cm within statistic uncertainties.

\subsection{Absorption Length}
\begin{figure}[htbp] \centering
  \includegraphics[width=\linewidth]{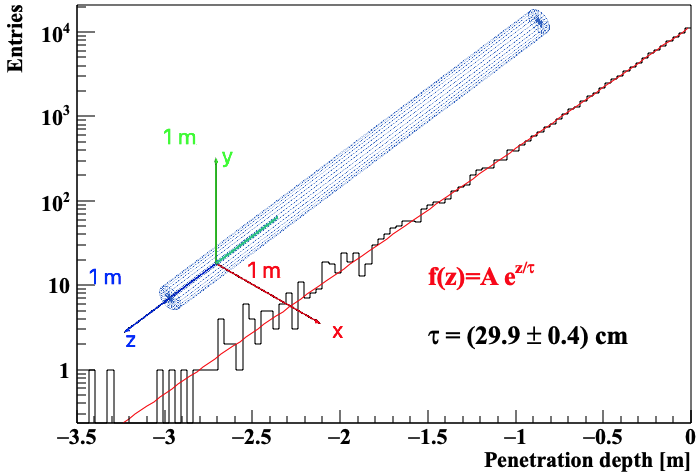}
  \caption{Absorption length in CsI crystal.}
  \label{f:absorption}
\end{figure}

In addition to be scattered, optical photons can also be absorbed by a medium. The absorption length is defined through the property \verb|AbsLength| in Geant4, as shown in Listing~\ref{l:csi}. Again, to verify the effectiveness of this configuration, 3.5~eV optical photons are shot along the z axis of a long rectangular CsI crystal as shown in the Geant4 visualization imbedded in Figure~\ref{f:absorption}. The distance between the origin and the absorption point of each trajectory is accumulated in the histogram shown in Figure~\ref{f:absorption}. It is nicely described by an exponetial function (red) and the constant $\tau=29.9\pm0.4$~cm given by the fitting matches the set value of 30 cm within statistic uncertainties.

Note that scattering parameters are disabled in this simulation to avoid interference, and vice versa in the previous Rayleigh scattering simulation.

\section{Surface Optical Properties} \label{s:sur}

When an optical photon generated in a dielectric material travels to the boundary of that material, it reaches an optical interface (or optical surface using Geant4 terminology) between two materials, the one that it is about to leave and the one that it is about to enter. Optical properties of the material where the photon is generated cannot determine the fate of the photon on its own. As a minimal requirement, optical properties of the material that the photon is about to enter also need to be specified. More generally, certain properties of the interface itself may need to be specified as well, such as its roughness, etc., and they are independent from optical properties of the two materials on the two sides of the interface. The \verb|:surf| tag is dedicated to the configuration of these surface optical properties. If the \verb|:surf| tag is not specified, the program defaults to the ideal surface mentioned in the next section.

\subsection{Ideal Surface}\label{s:ideal}
An ideal interface is the simplest case of an optical interface. In this case, no additional property of the interface itself is needed. The fate of an optical photon reaching the interface is determined solely by the refractive indexes of the two materials on the two sides of the interface, $n_1$ and $n_2$, following Snell's law and Fresnel equations~\cite{g4doc}.

\begin{figure}[htbp] \centering
  \includegraphics[trim={0 0 0 2cm}, clip, width=0.85\linewidth]{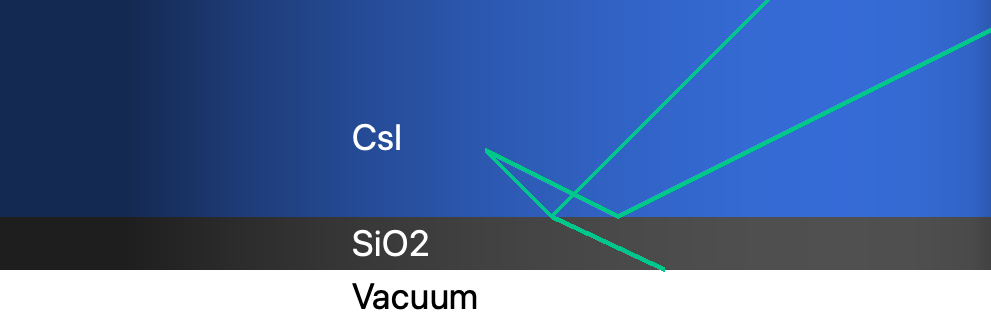}
  \caption{Optical photons (green) shot from the center of CsI on an ideal optical interface between CsI and a PMT window (SiO$_2$).}
  \label{f:ideal}
\end{figure}

Figure~\ref{f:ideal} visualizes the behavior of optical photons (green) on an ideal optical interface between CsI and a PMT window made of SiO$_2$. Photons are shot from the CsI crystal (blue) towards the PMT window (black) at different incident angles. As expected, at a smaller incident angle, some photons are refracted into the PMT window while others are reflected back into the CsI crystal, at a larger angle, all photons are totally reflected, following Snell's law and Fresnel equations.

This simulation is done by placing a CsI crystal defined in Listing~\ref{l:det} and a PMT window defined in Listing~\ref{l:pmt} next to each other in a detector definition file. The refractive indices of the two materials are provided in the included files \verb|CsI77K.tg| (Listing~\ref{l:csi}) and \verb|SiO2.tg| (Figure~\ref{f:nwin}), respectively. No more optical property is needed to define the interface between the two materials. A Geant4 macro file similar to Listing~\ref{l:run} is used to shoot 3.5~eV optical photons from the CsI crystal towards the PMT window at different incident angles.

\subsection{The UNIFIED Model}

Defining optical properties of a surface goes beyond the ideal interface described above. Geant4 provides several models to define more complex optical surfaces, among which the \texttt{UNIFIED} model is the most versatile one~\cite{g4doc}. Listing~\ref{l:surf} shows an example of defining an optical surface using the \texttt{UNIFIED} model in the extended plain text geometry syntax.  In the following sections, we first demonstrate the implementation of the \texttt{UNIFIED} model using the extended syntax for the \verb|dielectric_dielectric| interface with various surface finishes. The \verb|dielectric_metal| interface will be discussed after that.

\subsection{Unpainted Dielectric Surfaces}
While Geant4 does not explicitly use the term \emph{unpainted} as a reserved keyword, it is used here to categorize surfaces where the optical boundary is defined solely by the interface between two dielectric media. Two finishes are available for these surfaces in the UNIFIED model: $\verb|polished|$ and $\verb|ground|$, each introduces more optical parameters for increased flexibility.

\subsubsection{Polished Finish}

The polished finish in the UNIFIED model can be regarded as an extension of the ideal interface discussed in Section \ref{s:ideal}. It allows users to specify two new parameters: Reflectivity (R) and Transmittance (T). Both parameters must be within the range $[0, 1]$, and their sum must satisfy $0 \le R+T \le 1$. If $R+T>1$, Geant4 will redefine $R$ as $R/(R+T)$ and $T$ as $T/(R+T)$ such that the sum of the newly defined $R$ and $T$ is equal to 1.

It is crucial to note that these parameters are settings, not physically calculated probabilities derived from Snell's law or Fresnel equations.

Specifically, Transmittance (T) defines the ratio of incident photons that immediately pass through the surface without further optical processing. Photons transmitted in this manner retain the polarization state of the incident photon. Setting $T > 0$ may seem artificial, but it is particularly useful for modeling complex physical components as simplified geometric planes. A primary example is the metal wire mesh commonly found in time projection chambers. Rather than simulating each individual wire, a user can define a flat plane with $T$ equal to the open-area ratio of the mesh, allowing transmitted photons to continue their tracks in the subsequent volume. Another typical use case is a partial polarizer, where a specific fraction of light is intended to pass through the boundary while strictly preserving its original polarization state.

Conversely, Reflectivity (R) specifies the ratio of photons reserved for processing by physics laws; they will be subjected to the standard Geant4 reflection/refraction calculations based on the refractive indices and the angle of incidence.

A third parameter, Absorption (A), is not specified by the user but is instead calculated from the user-defined settings via the simple energy conservation relation:
$$A = 1 - T - R$$
This parameter represents the ratio of incident photons that are absorbed at the surface.

The ideal interface described in Section~\ref{s:ideal} represents a specific configuration of the polished finish where $R=1$ (and consequently $T=0, A=0$). Since these values constitute the model's default settings, a surface declared as $\texttt{polished}$ without explicitly defined $R$ or $T$ values will behave identically to an ideal interface.

\begin{figure}[htbp]\centering
  \includegraphics[width=0.9\linewidth]{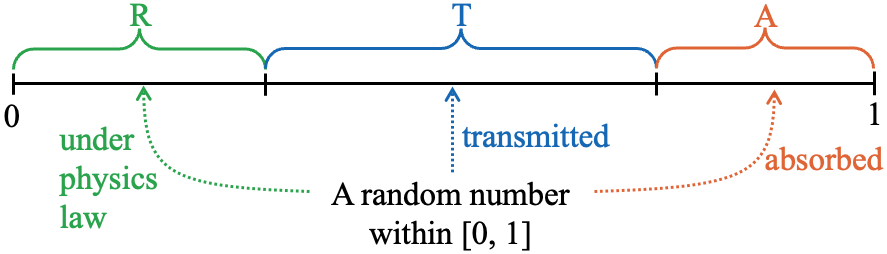}
  \caption{The partitioning of the random number range $[0, 1)$ by the $R$, $T$, and $A$ parameters determines the fate of a photon incident on the surface.}
  \label{f:rta}
\end{figure}

In a more general configuration, where $R>0$, $T>0$, and $A>0$ such that $R+T+A=1$, the fate of an incident photon is determined by a pseudo-random number generator, as illustrated conceptually in Figure~\ref{f:rta}. When a photon reaches the surface, a uniformly distributed random number $u$ is generated within the range $[0, 1)$. The result is determined by comparing $u$ against the defined parameters:
\begin{enumerate}
  \item If $u \in [0, R)$, the photon's outcome is calculated based on the standard optical physics rules, including Snell's Law and Fresnel equations.
  \item If $u \in [R, R+T)$, the photon is directly transmitted through the interface.
  \item If $u \in [R+T, 1)$, the photon is absorbed at the surface.
\end{enumerate}

It is worth noting that the parameter names are somewhat misleading due to the probabilistic nature of the first case. Even when a random number falls into the \emph{Reflectivity} range (Case 1), the subsequent physics-based calculation may still result in the photon being refracted through the interface, provided it is allowed by the refractive indices and the angle of incidence. Thus, \emph{Reflectivity} ($R$) does not represent the probability of reflection, but rather the probability of applying a physics-based calculation. Similarly, \emph{Transmittance} ($T$) is not the probability of transmission calculated from physics, but an artificial setting that controls the probability of a photon passing through the interface without any change in direction or polarization.

Listing~\ref{lst:polished} demonstrates the implementation of this finish in GEARS using the extended text geometry syntax. Here, a polished interface named \texttt{CsI2Epoxy} is defined between a CsI crystal (copy number 1) and an optical epoxy layer (copy number 2).

\begin{lstlisting}[caption=Polished surface definition for CsI-Epoxy interface., label=lst:polished]
:surf CsI2Epoxy CsI:1 Epoxy:2
  type dielectric_dielectric
  model unified
  finish polished
  property
    photon_energies 2 2.5*eV 5.0*eV
       Reflectivity   0.5    0.5
      Transmittance   0.5    0.5
\end{lstlisting}

\begin{figure}[htbp]
  \centering
  \includegraphics[trim={1cm 0 0 6cm}, clip, width=0.8\linewidth]{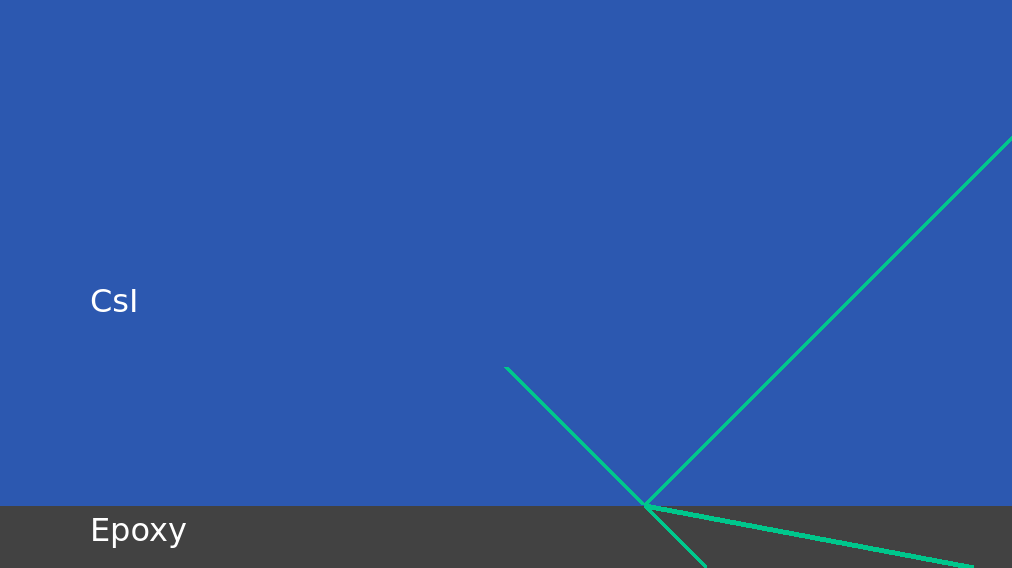}
  \caption{Optical photons (green) shot from the center of CsI on a \texttt{polished} CsI--Epoxy interface with $R=0.5$ and $T=0.5$.}
  \label{f:polished05}
\end{figure}

Figure~\ref{f:polished05} demonstrates the result with this setting. Because the sum of probabilities is unity ($R+T=1$), no photons are absorbed. The large $T$ is not set based on reality, but to exaggerate its effect. Half of the photons undergo transmission, passing straight through the interface without any change in direction. The other half undergo a standard physical calculation, resulting in either reflection or refraction.

This distinction between the transmitted and refracted photons is clearly visible in the figure: the transmitted ones continue along their original trajectory, whereas the refracted ones bend further at an angle relative to the surface normal, consistent with the change in refractive index between the two media.

\begin{figure}[htbp]
  \centering
  \includegraphics[trim={0 0 0 5cm}, clip, width=0.8\linewidth]{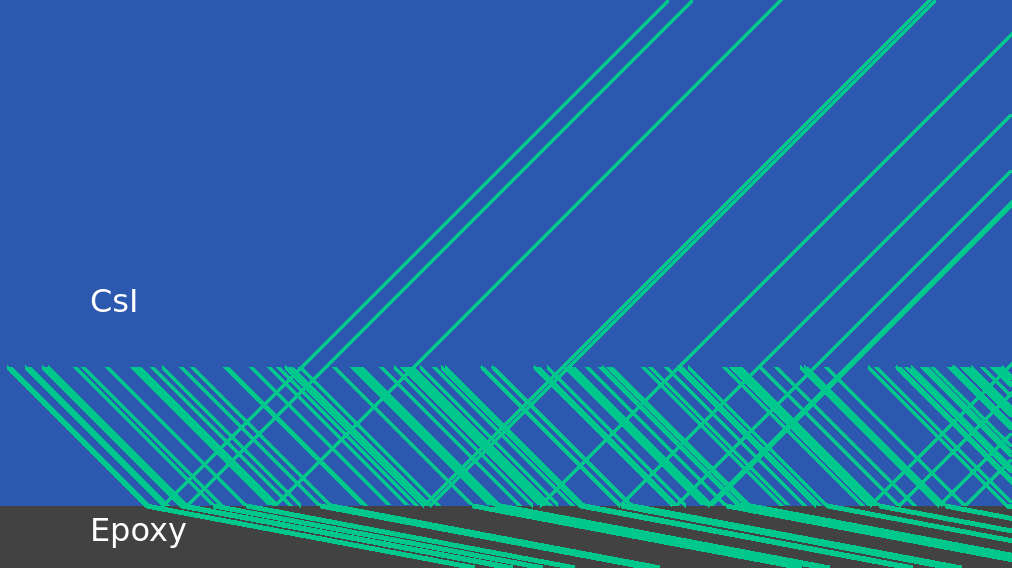}
  \caption{Optical photons (green) shot from the middle plane of CsI on a polished CsI-Epoxy interface with $R=0.5$ and $T=0$.}
  \label{f:polished05new}
\end{figure}

Figure~\ref{f:polished05new} demonstrates another configuration featuring a polished interface with $R=0.5$ and $T=0$. In this scenario, half of the incident photons are absorbed at the boundary, while the other half undergo a calculation via physics laws. To illustrate this, photons are directed onto the interface at an oblique angle. Approximately half of the tracks terminate at the interface, verifying the $A=1-R-T=0.5$ setting. The remaining photons either refract into the epoxy or reflect specularly, consistent with physics laws.

\subsubsection{Ground Finish}

\begin{figure}[htbp]
  \centering
  \begin{tikzpicture}
    % Shaded bulk dielectric material
    \fill[cyan!30] (0,-1) -- (0,0) -- (1.5,-0.2) -- (2.5,0.4) -- (3.5,-0.3) -- (4.5,0.2) -- (5.5,-0.1) -- (6.5,0.3) -- (6.5,-1) -- cycle;

    % Jagged micro-facet boundary
    \draw (0,0) -- (1.5,-0.2) -- (2.5,0.4) -- (3.5,-0.3) -- (4.5,0.2) -- (5.5,-0.1) -- (6.5,0.3);

    % Average surface line
    \draw[dashed, blue] (0,0.05) node[above] {Average Surface} -- (6.5,0.05);

    % Define coordinates for the highlighted micro-facet
    \coordinate (Center) at (3.0, 0.05);

    % Draw Average Surface Normal
    \draw[->, >=stealth] (Center) -- ++(0, 1.5) coordinate (AvgNorm)
    node[above] {Average Surface Normal};

    % Draw Micro-facet Normal
    % Vector P1(2.5, 0.4) to P2(3.5, -0.3) is (1.0, -0.7).
    % Perpendicular normal vector is (0.7, 1.0). Length = sqrt(0.49+1) = 1.22.
    % Scaled for drawing: (0.7/1.22, 1/1.22) * 1.5 = (0.86, 1.23)
    \draw[->, >=stealth, red] (Center) -- ++(0.86, 1.23) coordinate (FacetNorm)
    node[above right] {Micro-facet Normal};

    % Angle alpha arc (Corrected Order)
    \pic[draw, ->, >=stealth, "$\alpha$", angle eccentricity=1.2, angle radius=1cm] {angle = FacetNorm--Center--AvgNorm};

    \node at (0.8, 1.3) {Dielectric 1};
    \node at (0.8,-0.8) {Dielectric 2};
  \end{tikzpicture}
  \caption{2D schematic illustrating the local micro-facet normal relative to the average surface normal by an angle $\alpha\in[0,90]$ degrees.}
  \label{f:facets}
\end{figure}
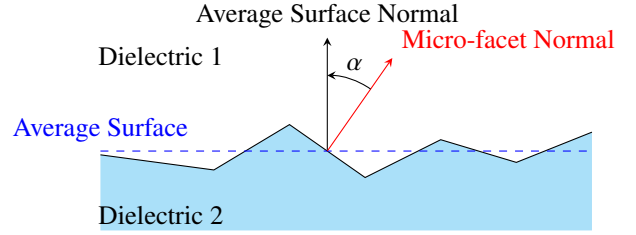

The \texttt{ground} finish within the \texttt{UNIFIED} model is designed to simulate a rough interface between two dielectric materials. The roughness is modeled by treating the interface as a collection of microscopic facets, or micro-facets in short, randomly oriented around the average surface normal, as shown in Figure~\ref{f:facets}.

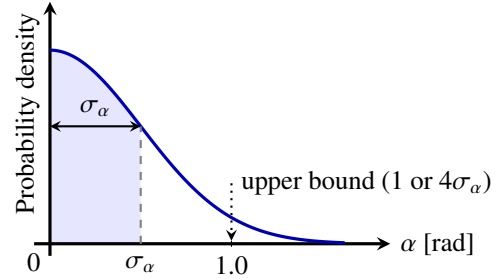
\begin{figure}[htbp]
  \centering
  \begin{tikzpicture}[xscale=3, yscale=0.8]
    \def\sig{0.4}   % sigma_alpha in radians
    \def\Ag{3.2}    % peak height

    % --- Shaded region for 0 to sigma ---
    \fill[blue!10, domain=0:\sig, samples=50]
    (0, 0) -- plot (\x, {\Ag*exp(-0.5*(\x/\sig)^2)}) -- (\sig, 0) -- cycle;

    % --- Axes ---
    \draw[->, >=stealth, very thick] (-0.07, 0) -- (1.5, 0)
    node[right] {$\alpha$ [rad]};
    \draw[->, >=stealth, very thick] (0, -0.2) -- (0, 4.0)
    node[above left, rotate=90] {Probability density};

    % --- Gaussian curve (domain 0 to ~3 sigma) ---
    \draw[very thick, blue!70!black, domain=0:1.3, samples=100, smooth]
    plot (\x, {\Ag*exp(-0.5*(\x/\sig)^2)});

    % --- Tick marks (only 1.0; skip 0.5 because sigma_alpha marks it, skip 1.5 because pi/2 is nearby) ---
    \draw (0.8, 0.05) -- (0.8, -0.05) node[below] {$1.0$};
    \node[below left] at (0, 0) {$0$};

    % --- pi/2 boundary ---
    \draw[->, >=stealth,dotted, thick] (0.8, 1) node[above, right] {upper bound (1 or $4\sigma_\alpha$)} -- (0.8, 0.05);

    % --- Sigma_alpha marker ---
    \draw[dashed, thick, gray] (\sig, 0) -- (\sig, {\Ag*exp(-0.5)});
    \node[below] at (\sig, -0.05) {$\sigma_\alpha$};

    % --- Sigma_alpha bracket arrow ---
    \draw[<->, >=stealth, thick] (0, {\Ag*exp(-0.5)}) -- (\sig, {\Ag*exp(-0.5)})
    node[midway, above] {$\sigma_\alpha$};
  \end{tikzpicture}
  \caption{Gaussian distribution of the micro-facet normal's polar angle $\alpha$, with a standard deviation of $\sigma_\alpha$, which characterizes the surface roughness.}
  \label{f:sa}
\end{figure}

The azimuthal angle $\phi$ of the micro-facet normal with respect to the average surface normal is uniformly sampled within $[0, 2\pi)$, which cannot be seen in the 2D schematic. The polar angle $\alpha\in[0,\pi/2]$ between the micro-facet normal and the average surface normal is sampled from a Gaussian distribution with mean 0 and standard deviation $\sigma_\alpha$, as shown in Figure~\ref{f:sa}. If $\alpha$ is larger than $\pi/2$, an incident photon from the upper dielectric material will hit the back side of a micro-facet, which is physically impossible.

Apparently, the larger $\sigma_\alpha$ is, the rougher the surface is. Empirically, lapped or machine-polished surfaces generally correspond to $\sigma_\alpha \le 0.05$ radian, while satin or fine-ground finishes typically fall within the 0.09--0.21 radian range. Coarser surfaces, such as those that are milled or sandblasted, may require values exceeding 0.26 radian. When $\sigma_\alpha$ is too large, say, 1.0 radian, the chance for a randomly sampled $\alpha$ to be larger than $\pi/2$ becomes significant, a few new samplings will be done until a valid $\alpha$ is obtained. This is highly inefficient and should be avoided. A new parameter, \texttt{sigma\_alpha}, is introduced to specify $\sigma_\alpha$, and its upper bound is set to the smaller of $1$ and $4\sigma_\alpha$ to ensure sampling efficiency.

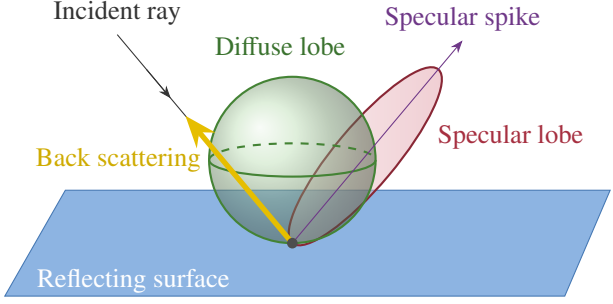
\begin{figure}[htbp]
  \centering
  \begin{tikzpicture}[>=Stealth]
    \definecolor{surfblue}{RGB}{80,140,210}
    \definecolor{lobegreen}{RGB}{90,175,70}
    \definecolor{lobeyellow}{RGB}{230,190,0}
    \definecolor{lobered}{RGB}{195,55,75}
    \definecolor{spikecolor}{RGB}{110,55,135}
    % Reflecting surface
    \fill[surfblue!65, draw=surfblue!85!black, line width=0.4pt]
    (-3.8,-0.6) -- (3.6,-0.6) -- (4.2,0.8) -- (-3,0.8) -- cycle;
    % Incidence point
    \coordinate (O) at (0,0.1);
    % Specular lobe

    \begin{scope}[rotate around={-40:(O)}]
      \shade[left color=lobered!18, right color=lobered!55, opacity=0.45]
      ([shift={(0,1.5)}]O) ellipse [x radius=0.4, y radius=1.5];
      \draw[lobered!70!black, thick]
      ([shift={(0,1.5)}]O) ellipse [x radius=0.4, y radius=1.5];
    \end{scope}

    % Diffuse lobe (full sphere sitting on the surface)
    \shade[ball color=lobegreen!50!white, opacity=0.50]
    (0,1.2) circle (1.1);
    \draw[lobegreen!75!black, thick]
    (0,1.2) circle (1.1);
    % Equator ellipse
    \draw[lobegreen!75!black, thick]
    (-1.1,1.2) arc[start angle=180, end angle=360, x radius=1.1, y radius=0.22];
    \draw[lobegreen!75!black, thick, dashed]
    (-1.1,1.2) arc[start angle=180, end angle=0, x radius=1.1, y radius=0.22];
    \node[lobegreen!65!black] at (-0.16,2.7) {Diffuse lobe};
    % Specular lobe label
    \node[lobered!85!black, anchor=west] at (1.8,1.5) {Specular lobe};
    % Specular spike
    \draw[->, spikecolor] (O) -- ({3.5*sin(40)},{0.1+3.5*cos(40)}) node[above] {Specular spike};
    % Incident ray
    \draw[black!80] ({-3.6*sin(40)},{0.1+3.6*cos(40)}) node[above] {Incident ray} -- (O);
    \draw[->, black!80] ({-3.0*sin(40)},{0.1+3.0*cos(40)}) -- ({-2.5*sin(40)},{0.1+2.5*cos(40)});
    % Backscatter arrow
    \draw[->, lobeyellow, line width=2pt] (O) -- ({-2.2*sin(40)},{0.1+2.2*cos(40)});
    \node[lobeyellow!90!black] at (-2.3, 1.2) {Back scattering};
    \fill[black!70] (O) circle (2pt);
    \node[white] at (-2.1,-0.4) {Reflecting surface};
  \end{tikzpicture}
  \caption{Four different reflection scenarios from a rough surface.}
  \label{f:4rs}
\end{figure}

On highly rough surfaces, an optical photon may undergo multiple reflections across several micro-facets. In such cases, Geant4 determines the refraction angle, and the reflection and refraction probabilities based on the first micro-facet orientation. However, the final reflection angle falls into four different scenarios as shown in Figure~\ref{f:4rs}. The probability of each scenario is determined by three additional parameters introduced below.

\begin{description}
  \item[Back scattering] happens when a photon undergoes multiple reflections within deep grooves on the surface and eventually scatters back toward its starting point. The probability of this scenario, $C_{bs} \in [0,1]$, can be specified by parameter \texttt{backscatterConstant}.
  \item[Specular spike] represents specular reflection about the average surface normal. Its probability $C_{ss} \in [0,1]$ can be specified by parameter \texttt{specularSpikeConstant}.
  \item[Specular lobe] represents a reflection about the micro-facet normal rather than the average surface normal. The probability of this scenario, $C_{sl} \in [0,1]$, can be specified by parameter \texttt{specularLobeConstant}.
  \item[Diffuse lobe] represents Lambertian reflection. The probability of this scenario is $C_l = 1 - (C_{bs} + C_{ss} + C_{sl})$. No new parameter is needed to specify this scenario.
\end{description}

The usage of these new parameters is demonstrated in Listing~\ref{l:cbs}, which shows the application of this finish to a nonpolished CsI crystal wrapped with Teflon.

\begin{lstlisting}[caption={Example ground surface definition for CsI-Teflon interface.}, label={l:cbs}, numbers=left, numbersep=5px, numberstyle=\color{blue}]
:surf CsI2Teflon CsI:1 Teflon:2
  type dielectric_dielectric
  model unified
  finish ground
  sigma_alpha 0.1
  property photon_energies 2 2.5*eV 5.0*eV
       backscatterConstant   0.1    0.2
     specularSpikeConstant   0.2    0.1
      specularLobeConstant   0.7    0.7
\end{lstlisting}

\begin{figure}[htbp]
  \centering
  \includegraphics[width=0.8\linewidth]{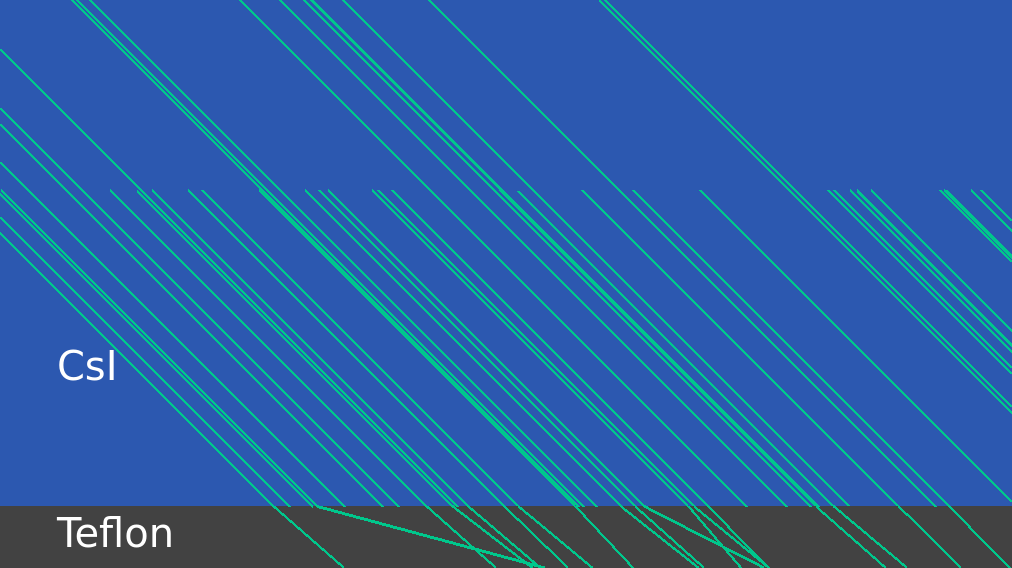}
  \caption{Optical photons (green) shot from the middle plane of CsI to the CsI-Teflon interface with ground finish, where $C_{bs}=1$ and $\sigma_\alpha=0.1$ radians. Starting points are covered by the backscattered trajectories.}
  \label{f:back}
\end{figure}

Figure~\ref{f:back} shows the case where $C_{bs}=1$ and other constants are zero. Within CsI, photons undergo backscattering reflection about the average surface normal. The directions of refracted photons in Teflon are randomized due to the small variation of the micro-facet orientation ($\sigma_\alpha=0.1$ radians).

\begin{figure}[htbp]
  \centering
  \includegraphics[trim={8cm 0 0 10cm}, clip, width=0.8\linewidth]{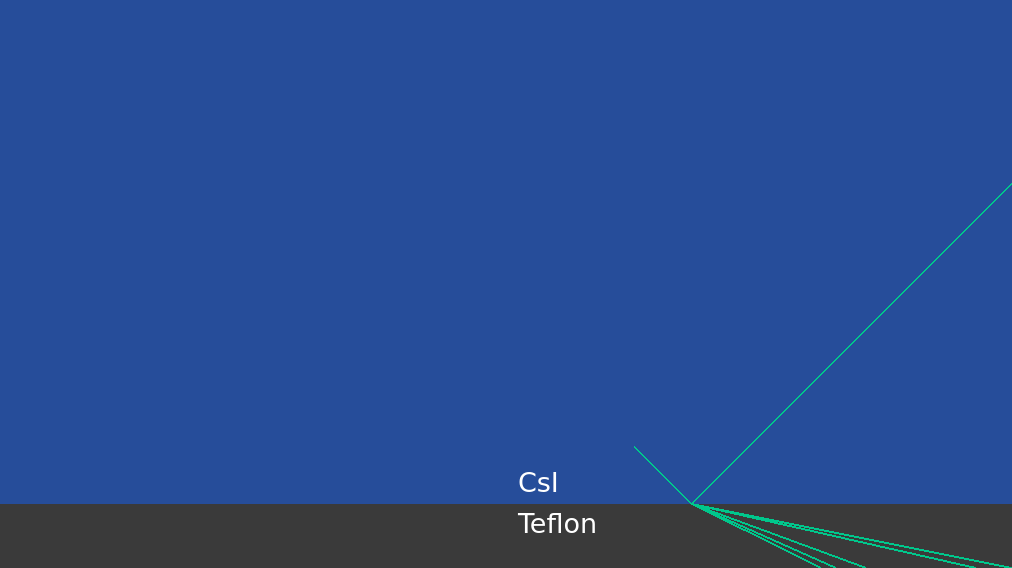}
  \caption{Optical photons (green) shot from the center of CsI to the CsI-Teflon interface with ground finish, where $C_{ss}=1$ and $\sigma_\alpha=0.1$ radians.}
  \label{f:css1}
\end{figure}

Figure~\ref{f:css1} shows the case where $C_{ss}=1$ and other constants are zero. Within CsI, photons undergo specular spike reflection about the average surface normal. The refracted photons in Teflon are diffused a bit due to the small variation of the micro-facet orientation ($\sigma_\alpha=0.1$ radians).

\begin{figure}[htbp]
  \centering
  \includegraphics[trim={4cm 0 2cm 10cm}, clip, width=0.8\linewidth]{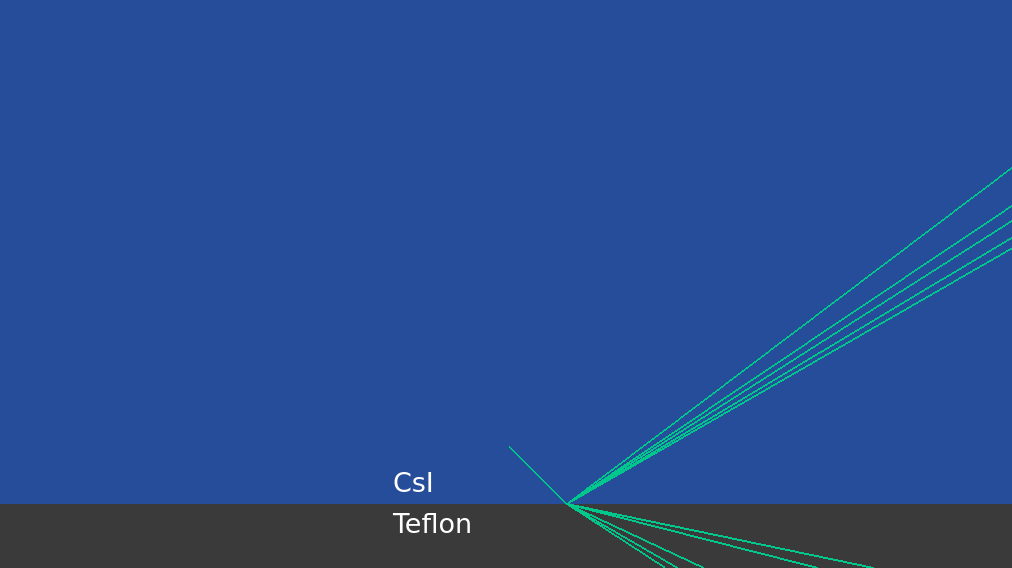}
  \caption{Optical photons (green) shot from the center of CsI to the CsI-Teflon interface with ground finish, where $C_{sl}=1$ and $\sigma_\alpha=0.1$ radians.}
  \label{f:csl1}
\end{figure}

Figure~\ref{f:csl1} shows the case where $C_{sl}=1$ and other constants are zero. Within CsI, photons undergo fuzzy specular reflection about the average surface normal. The directions of both reflected and refracted photons are randomized due to the small variation of the micro-facet orientation ($\sigma_\alpha=0.1$ radians).

If none of the constants is specified, they are initialized to zero. The finish defaults to pure Lambertian reflection because $C_l = 1-(C_{ss} + C_{sl} + C_{bs}) = 1$.

\begin{figure}[htbp]
  \centering
  \includegraphics[trim={1cm 0 0 6cm}, clip, width=0.8\linewidth]{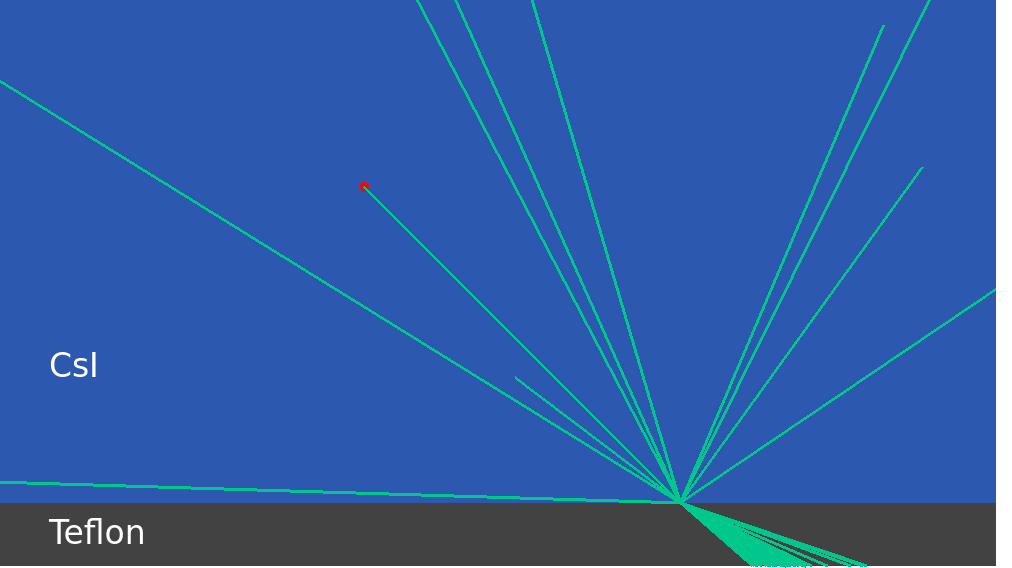}
  \caption{Optical photons (green) shot from the center of CsI to the CsI-Teflon interface with ground finish, where $C_{l}=1$ and $\sigma_\alpha=0.1$ radians.}
  \label{f:cl1}
\end{figure}

Figure~\ref{f:cl1} is generated using settings in Listing~\ref{l:cbs} without line 6, 7, 8, 9. Photons undergo Lambertian reflection in CsI.  Photons refracted in Teflon are diffused a bit due to the small variation of the micro-facet orientation ($\sigma_\alpha=0.1$ radians).

\begin{figure}[htbp]
  \centering
  \begin{tikzpicture}
    \definecolor{blue}{RGB}{50, 108, 210}

    \filldraw[fill=blue,    rounded corners] (-3.0,-2.4) rectangle (3.0, 2.0);
    \filldraw[fill=blue!85, rounded corners] (-2.2,-2.2) rectangle (2.2, 0.5);
    \filldraw[fill=blue!65, rounded corners] (-1.6,-2.0) rectangle (1.6,-1.0);

    \node[text=white] (G) at (0, 1.5) {finish: \texttt{ground}};
    \node[text=white] (P) at (0, 0) {finish: \texttt{polished}};
    \node[text=black] (I) at (0,-1.5) {ideal interface};

    \draw[->, white] (G) to node{$\sigma_\alpha = 0$, $C_{ss} = 1$} (P);
    \draw[->, white] (P) to node{$R=1$, $T=0$} (I);
  \end{tikzpicture}
  \caption{Venn diagram illustrating the hierarchical relationship between different finishes. The arrows indicate how applying specific constraints (e.g., $R=1, T=0$) transitions the functional behavior from the superset \texttt{ground} finish down to the \texttt{polished} finish, and finally the ideal interface.}
  \label{f:venn}
\end{figure}
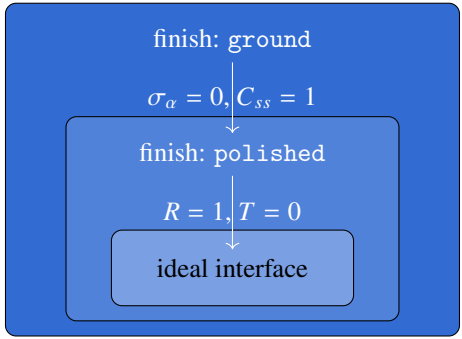

It is worth noting that the \texttt{ground} finish serves as a functional superset of the \texttt{polished} finish and the ideal interface. As shown in Figure \ref{f:venn}, by setting $\sigma_\alpha = 0$ and $C_{ss} = 1$, the standard polished logic is exactly replicated. Under these constraints, if we further require $R=1$, all incident photons will propergate according to physics laws, and the ideal optical interface logic is recovered.

\subsection{Painted Dielectric Surfaces}

\begin{figure}[htbp]
  \centering
  \begin{tikzpicture}[>=stealth]
    \fill[black!70](-4,-0.6) rectangle (4,0.8);
    \node[text=white, align=center] at (0, 0) {Painted substrate (e.g., floor)};
    \fill[teal!70](-4,0.8) rectangle (4,1.4);
    \node[text=white, align=center] at (0, 1.1) {Paint (e.g., wax)};

    \node at (0, 2.6) {Dielectric (e.g. air)};
    \draw[->, thick] (-2, 2.6) -- (0, 1.4) node[midway, left, xshift=-8pt] {Incident};
    \draw[->, thick] (0, 1.4) -- (2, 2.6) node[midway, right, xshift=8pt] {Reflected};
  \end{tikzpicture}
  \caption{Schematic of \texttt{...frontpainted} finishes. Photons hit the paint in front of the painted substrate.}
  \label{f:frontp}
\end{figure}
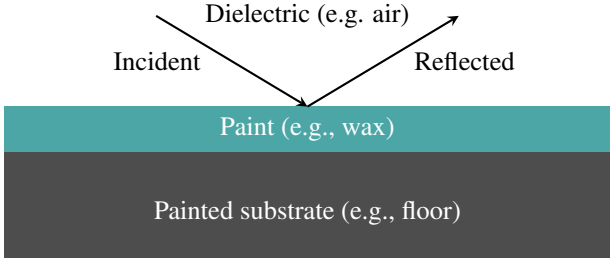

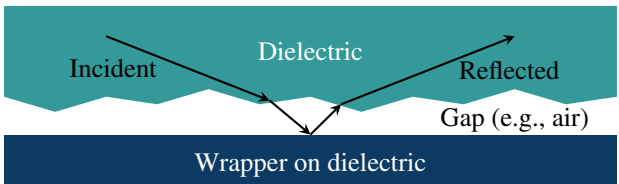
\begin{figure}[htbp]
  \centering
  \begin{tikzpicture}[xscale=1.35, >=stealth, thick,
      define color/.code={\definecolor{darkblue}{RGB}{12, 58, 99}
        \definecolor{midblue}{RGB}{50, 108, 163}
      \definecolor{photongreen}{RGB}{50, 205, 50}},
      define color,
    ]

    \fill[fill=teal!80] (0,2.0) -- (0,3.2) -- (6,3.2) -- (6,2.0)
    -- (5.5, 1.9) -- (5.0, 2.1) -- (4.5, 1.9) -- (4.0, 2.0)
    -- (3.5, 1.8) -- (3.0, 2.0) -- (2.5, 1.9) -- (2.0, 2.1)
    -- (1.5, 1.9) -- (1.0, 2.0) -- (0.5, 1.8) -- cycle;

    \node[white] at (3, 2.6) {Dielectric};

    % Air Gap (Label lowered into tighter gap)
    \node at (5.0, 1.7) {Gap (e.g., air)};

    % Paint Layer (Kept at same height, making the gap visually tighter)
    \fill[fill=darkblue] (0,0.8) rectangle (6,1.5);
    \node[white] at (3, 1.1) {Wrapper on dielectric};

    \draw[->] (1, 2.8) -- (2.6, 1.95) node[midway, left, xshift=-8pt] {Incident};
    \draw[->] (2.6, 1.95) -- (3.0, 1.5);
    \draw[->] (3.0, 1.5) -- (3.3, 1.9);
    \draw[->] (3.3, 1.9) -- (5, 2.8) node[midway, right, xshift=8pt] {Reflected};
  \end{tikzpicture}
  \caption{Schematic of \texttt{...backpainted} finishes. The wrapper is on the back side of the gap.}
  \label{f:backp}
\end{figure}

For surfaces painted or wrapped with reflective layers, the nomenclature in the \texttt{UNIFIED} model shifts: while \texttt{polished} and \texttt{ground} previously referred to the nature of the dielectric interface itself, here they characterize the paint. Explicitly, the descriptor \texttt{ground} indicates that the paint functions as a diffusive (Lambertian) reflector, whereas \texttt{polished} identifies the paint as a specular (mirror-like) reflector. The prefixes, \texttt{front} and \texttt{back}, distinguish the interaction sequence: \texttt{front} implies immediate photon-paint interaction (Figure~\ref{f:frontp}), whereas \texttt{back} implies the photon first traverses the gap before encountering the paint (Figure~\ref{f:backp}).

We will start with the discussion of the \texttt{...frontpainted} finishes and then proceed to the \texttt{...backpainted} ones.

\subsubsection{PolishedFrontPainted Finish}

The \texttt{polishedfrontpainted} finish defines a surface coated with a specularly reflective paint.

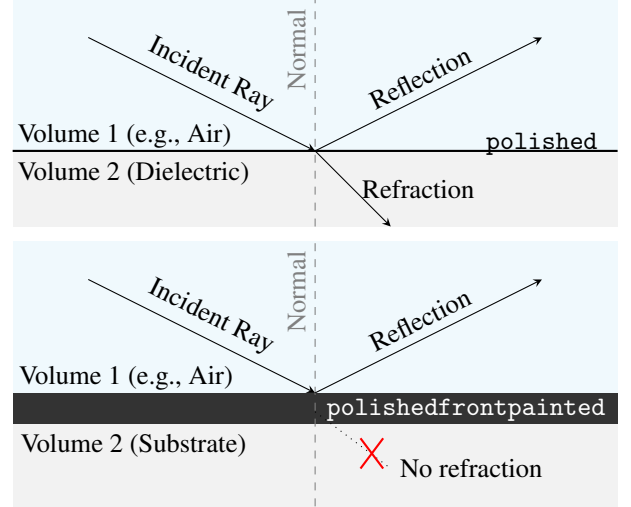
\begin{figure}[htbp]
  \centering
  \begin{tikzpicture}[>=stealth]
    \begin{scope}[shift={(0,0)}] % top figure: polished
      \fill[cyan!5] (-4, 0) rectangle (4, 2);
      \node at (-2.52, 0.2) {Volume 1 (e.g., Air)};

      \fill[gray!10] (-4, -1) rectangle (4, 0);
      \node at (-2.4, -0.3) {Volume 2 (Dielectric)};

      \draw[thick] (-4, 0) -- (4, 0);
      \node at (3, 0.1) {\texttt{polished}};

      \draw[dashed, gray] (0, -1) -- (0, 2) node[above left, rotate=90] {Normal};
      \draw[->] (-3, 1.5) -- (0, 0) node[midway, above, sloped] {Incident Ray};
      \draw[->] (0, 0) -- (3, 1.5) node[midway, above , sloped] {Reflection};
      \draw[->] (0, 0) -- (1, -1) node[midway, above, right] {Refraction};
    \end{scope}

    \begin{scope}[shift={(0,-3.2)}] % bottom figure: polished front painted
      \fill[cyan!5] (-4, 0) rectangle (4, 2);
      \node at (-2.5, 0.2) {Volume 1 (e.g., Air)};

      \fill[gray!10] (-4, -1.5) rectangle (4, 0);
      \node at (-2.4, -0.7) {Volume 2 (Substrate)};

      \fill[black!80] (-4, 0) rectangle (4, -0.4);
      \node[white] at (2, -0.22) {\texttt{polishedfrontpainted}};

      \draw[dashed, gray] (0, -1.5) -- (0, 2) node[above left, rotate=90] {Normal};

      \draw[->] (-3, 1.5) -- (0, 0) node[midway, above, sloped] {Incident Ray};
      \draw[->] (0, 0) -- (3, 1.5) node[midway, above, sloped] {Reflection};
      \draw[dotted] (0, -0.25) -- (1, -1) node[above, right] {No refraction};

      \draw[thick, red] (0.6, -0.6) -- (0.9, -1.0);
      \draw[thick, red] (0.6, -1.0) -- (0.9, -0.6);
    \end{scope}

  \end{tikzpicture}
  \caption{Comparison between the \texttt{polished} and \texttt{polishedfrontpainted} finishes. No refraction occurs in a  \texttt{polishedfrontpainted} surface.}
  \label{f:newpolished}
\end{figure}

The functional distinction between this and the \texttt{polished} finish lies in the interpretation of the reflectivity parameter, $R$. For a \texttt{polished} interface, $R$ dictates the probability that a photon undergoes an optical physics calculation to determine whether it reflects or refracts. Conversely, for a \texttt{polishedfrontpainted} finish, physics laws are bypassed. $R$ directly specifies the probability that a photon undergoes purely specular reflection. No refraction is possible for the \texttt{polishedfrontpainted} finish as shown in Figure~\ref{f:newpolished}. Parameter $T$ and the calculated absorptivity $A=1-R-T$ behave the same in both finishes.

A real-life application of this finish is the simulation of a first-surface (or front-surface) mirror (FSM), such as those used in telescopes or laser optics. Unlike a standard household mirror (second-surface mirror) where light passes through glass before reflecting, a FSM has the reflective coating on the front side of the substrate. A photon traveling in air strikes the coating immediately and undergoes specular reflection off the coating, without reaching the substrate beneath it at all. As a result, the optical properties of the substrate beneath the coating do not matter and hence do not need to be specified.

\begin{figure}[htbp]
  \centering
  \includegraphics[width=0.8\linewidth]{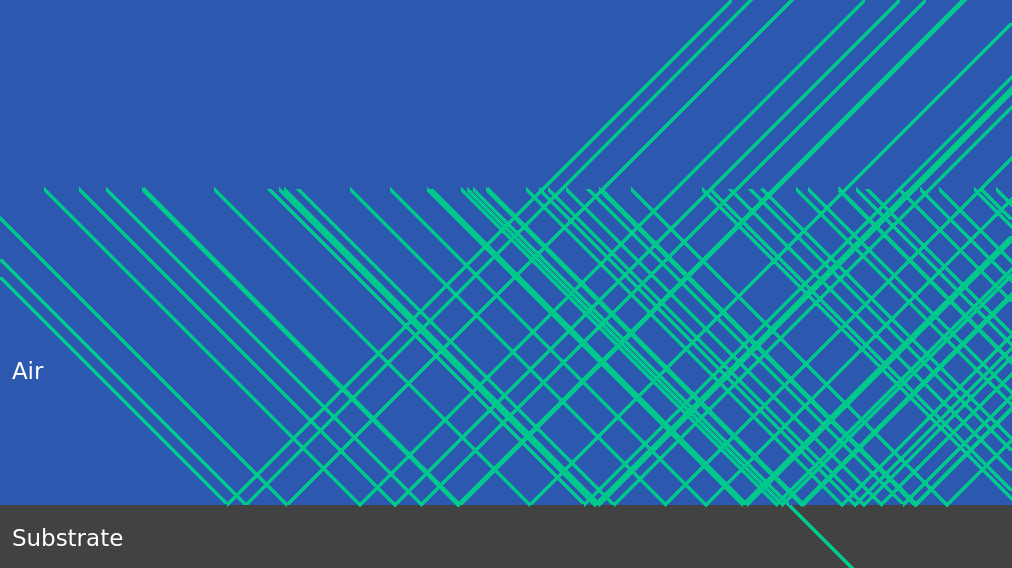}
  \caption{Optical photon trajectories (green) at a \texttt{polishedfrontpainted} interface. The surface acts as a first-surface mirror with 95\% reflectivity.}
  \label{f:mirror_sim}
\end{figure}

However, if a non-zero $T$ is given, that portion of photons will enter the subtrate as if the coating were not there. Users need to specify the optical properties of the substrate if they want to track these artificially transmitted photons. Listing~\ref{lst:mirror} gives an example of a FSM with 95\% reflectivity and 5\% transmittance. Its results are shown in Figure~\ref{f:mirror_sim}.

\begin{lstlisting}[caption={PolishedFrontPainted definition for a first-surface mirror.}, label={lst:mirror}]
:surf Air2Mirror Air:1 Mirror:1
  type dielectric_dielectric
  model unified
  finish polishedfrontpainted
  property photon_energies 2 2.5*eV 5.0*eV
              Reflectivity   0.95   0.95
              Transmittance  0.05   0.05
\end{lstlisting}

\subsubsection{GroundFrontPainted Finish}

This finish models a surface coated with a diffusive reflective layer. A canonical application is the integrating sphere, designed to homogenize incident light using multiple diffusive reflections. These devices typically consist of a hollow cavity, coated in its inner surface with a highly diffusive material such as white BaSO$_4$ powder or sintered PTFE (Teflon). In the context of luminous flux metrology, a light source is suspended centrally within the cavity to measure its total optical power. By capturing the full spherical emission ($4\pi$ steradians) and scattering it repeatedly against the cavity walls, this geometry spatially averages the light field, effectively eliminating errors that would otherwise arise from the source's specific angular directionality.

\begin{figure}[htbp]
  \centering
  \includegraphics[width=0.6\columnwidth]{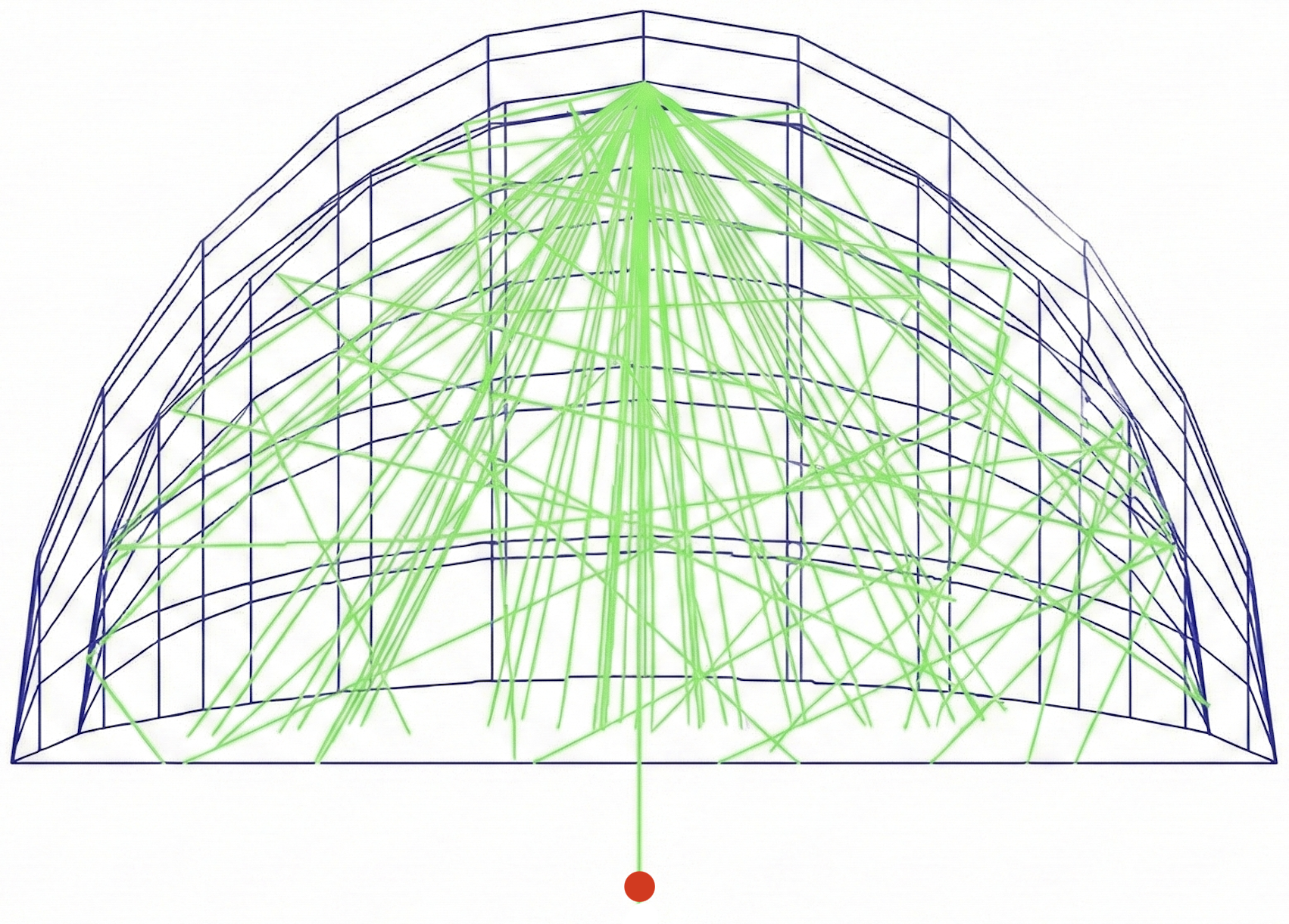}
  \caption{Optical photon trajectories (green) in the top half of an integrating sphere with a \texttt{groundfrontpainted} finish. Photons are shot upwards from the source (red dot).}
  \label{f:int_sphere_sim}
\end{figure}

Listing~\ref{lst:int_sphere} defines the optical surface for half of such a sphere, where the inner wall is treated as a 99\% reflective diffuse coating. Only 1\% of photons are absorbed on the coating. The output of the simulation is shown in Figure~\ref{f:int_sphere_sim}.

\begin{lstlisting}[caption={GroundFrontPainted definition for a Teflon-lined integrating sphere.}, label={lst:int_sphere}]
:surf Air2Teflon Air:1 Teflon:1
  type dielectric_dielectric
  model unified
  finish groundfrontpainted
  property photon_energies 2 2.5*eV 5.0*eV
              Reflectivity   0.99   0.99
             Transmittance   0      0
\end{lstlisting}

\subsubsection{Back Painted Finishes}
\begin{figure}[htbp]
  \centering
  \begin{tikzpicture}[>=stealth]
    \def\jagged{(-4, 1.6) -- (-2.8, 1.2) -- (-0.8, 1.5) -- (-0.4, 1.7) -- (0.64, 1.2) -- (1.6, 1.5) -- (2.8, 1.8) -- (4, 1.4)}

    \fill[cyan!15] (-4, 2.5) -- \jagged -- (4, 2.5) -- cycle;
    \node at (1.6, 2.2) {Dielectric 1};

    \node at (-2.7, 1) {Gap (Dielectric 2)};

    \fill[teal!85!black] (-4, 0.8) rectangle (4, 0.3);

    \node at (1.2, 1.4) {\emph{ground} finish};

    \node[rotate=-20] at (3.4, 1.8) {Micro};
    \node[rotate=-20] at (3.3, 1.45) {facets};

    \node[white] at (0, 0.55) {\emph{...frontPainted} finish};

    \draw[->, thick] (-3, 2) -- (-1.05, 1.45) node[midway,above,sloped] {Photon};
    \draw[->, thick] (-1.05, 1.45) -- (-0.2, 0.8);
    \draw[->, thick] (-0.2, 0.8) -- (0, 1.53);
    \draw[->, thick] (0, 1.53) -- (0.05, 2.3);
  \end{tikzpicture}
  \caption{The \texttt{...backpainted} finish as a combination of a \texttt{ground} finish on top of a \texttt{...frontpainted} finish.}
  \label{f:pbp}
\end{figure}
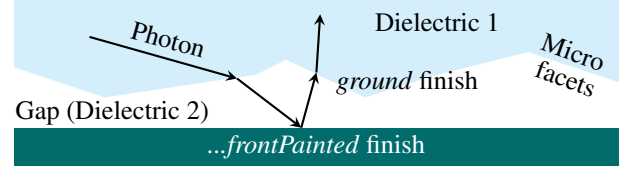

Back painted finishes can be understood as the combination of a \emph{ground} finish between dielectric 1 and the gap (dielectric 2) with a front-painted surface, as shown in the relation schematic in Figure~\ref{f:pbp}. Whether it is \emph{polished} or \emph{ground} is determined by the front painted finish, not the finish of the interface between dielectric 1 and the gap, which is always \emph{ground}:
$$groundBackPainted = ground + groundFrontPainted$$
$$polishedBackPainted = ground + polishedFrontPainted$$

It should be explicitly noted, however, that attempting this conceptual recreation requires the manual definition of three separate physical volumes: dielectric 1, the gap, and the external wrapper. The explicit backpainted finishes, therefore, provide a streamlined mechanism to simulate this complex three-tier volume interaction efficiently without requiring the manual definition of a microscopic gap volume. The refractive index of the gap must be explicitly defined as a parameter of the surface as shown in Listing~\ref{lst:pbp}. Otherwise, the photon will be killed upon encountering the surface.

\begin{lstlisting}[caption={Polished back painted definition for a plastic scintillator wrapped with mirror-reflective foil, where \emph{Rindex} specifies the refractive index of the gap, not that of the foil.}, label={lst:pbp}]
:surf scintilator2foil scintilator:1 foil:2
  type dielectric_dielectric
  model unified
  finish polishedBackPainted
  sigma_alpha  0.1
  property photon_energies 2 2.5*eV 5.0*eV
                    Rindex   1.0   1.0
\end{lstlisting}

\subsection{Summary of Dielectric-Dielectric Surfaces}
Table~\ref{t:dielectric_summary} summarizes parameters that determine the optical behavior of different types of \emph{dielectric-dielectric} surfaces. It is worth emphasizing that, while $n_{gap}$ is defined for the interface, $n_1$ and $n_2$ are properties of the bulk materials instead of the interface; they need to be defined so that photons can be properly propagated in those materials. From this point of view, a surface with a front painted finish requires zero parameter to function.

\begin{table}[htbp]
  \centering
  \caption{Summary of parameters that determine the optical behavior of different types of \emph{dielectric-dielectric} surfaces.}
  \label{t:dielectric_summary}
  \begin{tabular}{rll}
    \toprule
    {\small Surface and} & {\small Required}  & {\small Optional}\\
    {\small Finish Types} & {\small Parameters} & {\small Parameters}\\
    \midrule
    \emph{\small ideal} & $n_1, n_2$ & - \\
    \emph{\small polished} & $n_1, n_2$ & $R, T$ \\
    \emph{\small ground} & $n_1, n_2$ & $R, T, \sigma_{\alpha}, C_{bs}, C_{ss}, C_{sl}$ \\
    \emph{\small polishedFrontPainted} & $n_1$ & $R, T, n_2$ \\
    \emph{\small groundFrontPainted} & $n_1$ & $R, T, n_2$ \\
    \emph{\small polishedBackPainted} & $n_1, n_{gap}$ & $R, T, n_2, \sigma_{\alpha}, C_{bs}, C_{ss}, C_{sl}$ \\
    \emph{\small groundBackPainted} & $n_1, n_{gap}$ & $R, T, n_2, \sigma_{\alpha}, C_{bs}, C_{ss}, C_{sl}$ \\
    \bottomrule
  \end{tabular}
\end{table}

Default values of the optional parameters are $R=1, T=0, \sigma_{\alpha}=0, C_{bs}=C_{ss}=C_{sl}=0$, which results in the default behaviors listed in Table~\ref{t:default}.
\begin{table}[htbp]
  \centering
  \caption{Default behavior of \emph{dielectric-dielectric} surfaces. }
  \label{t:default}\small
  \begin{tabular}{rl}
    \toprule
    Finish Types & Default Behavior\\
    \midrule
    \emph{polished} & ideal reflection and refraction \\
    \emph{ground} &  Lambertian reflection, ideal refraction \\
    \emph{polishedFrontPainted} & specular reflection only \\
    \emph{groundFrontPainted} & Lambertian reflection only \\
    \emph{polishedBackPainted} & Lambertian and specular reflection \\
    \emph{groundBackPainted} & Lambertian reflection only \\
    \bottomrule
  \end{tabular}
\end{table}

\subsection{Dielectric-Metal Surfaces}
The interaction of electromagnetic waves with metallic surfaces is fundamentally different from dielectric-dielectric interfaces due to the high density of free electrons in conductors. When a photon strikes a metal, the oscillating electric field induces a rapid motion of these electrons, which in turn radiates an electromagnetic field that effectively cancels the incident field within the material. Consequently, electromagnetic waves cannot penetrate deep into the metal; instead, they are confined to a very thin skin depth, typically on the order of a few nanometers for visible and ultraviolet light. The macroscopic result of this interaction is high reflectivity.

For most metals used in detector construction, the reflectivity remains high until the photon energy approaches the material's plasma frequency ($\omega_p$). At this threshold, the electrons can no longer respond fast enough to cancel the field, and the metal begins to behave more like a transparent dielectric. The energy corresponding to the plasma frequency ($E_p = \hbar\omega_p$) typically falls in the vacuum ultraviolet (VUV) range for common metals. For example, aluminum has a high plasma energy of approximately $15.0$ eV ($\approx 83$ nm), making it an excellent reflector for most optical applications. Other common metals like silver and gold have plasma energies around 9.0 eV and 9.1 eV respectively, while copper sits at approximately 10.8 eV. Since the energy of scintillation photons is typically in the range of 2--5 eV, bulk metals can be regarded as opaque to them. For this reason, \emph{T} can be kept at its default value of 0, while \emph{R} can be set slightly lower than 1 to account for some surface absorption due to the energy loss to heat through ohmic dissipation. Cherenkov photons, on the other hand, may have energyies approaching or exceeding the plasma frequencies of the metal they interact with. \emph{T} can be adjusted according to experimental data to account for the increased transparency.

In addition to specifying the \emph{R} and \emph{T} parameters, Geant4 allows one to specify a complex refractive index, $\tilde{n} = n + ik$, where the imaginary part ($k$), also known as the extinction coefficient, dictates the rate of absorption within the skin depth. Parameters corresponding to these two indices are called \emph{realRindex} and \emph{imaginaryRindex}. It is important to note that these parameters are generally applicable for photon energies below the plasma frequency; above this threshold, the index of refraction for metals is no longer accurately described by the simple free electron model, requiring more complex treatment or empirical data.

\subsubsection{Polished}

The \emph{polished} \emph{dielectric\_metal} surface essentially exhibits the same behavior as the \emph{polishedFrontPainted} \emph{dielectric\_dielectric} surface. Its reflectivity and absorption rate can be fine-tuned by adjusting the \emph{R} and \emph{T} parameters guided by the general discussion provided previously.

\subsubsection{Ground}

The \emph{ground} \emph{dielectric\_metal} surface behaves much like the \emph{groundFrontPainted} \emph{dielectric\_dielectric} surface, in that the photon interacts immediately with the surface upon confronting it and there is no refraction. It differs notably in that if there is a reflection, the reflection type can be affected by the three user defined constants: $C_{ss}$, $C_{sl}$, and $C_{bs}$, as discussed in the \emph{ground} finish section for \emph{dielectric\_dielectric} interfaces.

\subsection{Look-Up Tables}

An alternative to the parameter-driven UNIFIED model is the Look-Up Table (LUT) approach. While the UNIFIED model relies on the manual specification of multiple optical properties, such as $\sigma_\alpha$, $C_{ss}$, $C_{sl}$, and $C_{bs}$, etc. LUT models, on the other hand, utilize pre-calculated or measured datasets to determine the fate of an incident photon. In these models, the complex interactions at the boundary are encapsulated within the table itself.

The primary advantage of the LUT approach is its ease of configuration. From a user's perspective, the setup is significantly more straightforward because it requires no manual parameter tuning. One simply selects a predefined finish, such as \emph{polishedVM2000Glue} or \emph{groundTyvekAir}, as shown in listing \ref{lst:lut_example}, and the simulation automatically handles the optical response. This turnkey functionality eliminates the need for the iterative optimization often required to find the correct $\sigma_\alpha$ or reflection coefficients for a specific physical surface, making it particularly useful for simulating standard detector materials that are well-characterized and consistent across different experimental setups.

\begin{lstlisting}[caption={Example of a LUT model configuration.}, label={lst:lut_example}]
:surf air2tyvek air:1 tyvek:2
  type dielectric_LUT
  model LUT
  finish groundTyvekAir
\end{lstlisting}

Despite their convenience, LUT models are fundamentally limited in scope compared to the UNIFIED model. Because they rely on specific experimental datasets, they are only available for a select range of surface finishes and material interfaces. Consequently, the LUT approach is not a universal solution; if a user is simulating a custom material, a unique surface treatment, or a boundary condition not covered by the existing libraries, they must revert to the UNIFIED model's parameter-based framework to manually construct the surface's optical profile.

\section{Conclusion}

In this work, we introduced an extension to the Geant4 plain text geometry syntax that allows for the comprehensive definition of optical properties without C++ programming. Through the newly developed \texttt{:prop} and \texttt{:surf} tags, users can seamlessly assign both constant and energy-dependent optical properties to bulk materials and intricate surface boundaries.

We validated this implementation against a broad spectrum of optical phenomena, successfully reproducing the expected physical behaviors for Cherenkov radiation, scintillation, Rayleigh scattering, and optical absorption. Additionally, we demonstrated the extension's robust capability to handle diverse surface configurations by extensively exploring the UNIFIED model, covering polished, ground, and painted finishes for both dielectric-dielectric and dielectric-metal interfaces.

By interpreting text-based optical property definitions at runtime, this approach bypasses the time-consuming C++ compilation cycle and makes Geant4 optical simulations significantly more accessible. The modular nature of the text files encourages the reuse of optical property definitions across different detector geometries. Ultimately, this development streamlines the iterative design of complex optical systems, facilitating rapid prototyping and advanced simulation efforts for future applications like the operation of pure CsI detectors at cryogenic temperatures in dark matter and neutrino experiments.

\section{Acknowledgements}
This work is supported by the NSF award OIA-2437416, PHY-2411825, and the Office of Research at the University of South Dakota. Computations supporting this project were performed on High Performance Computing systems at the University of South Dakota, funded by NSF award OAC-1626516.

\section{Declaration of Use of AI Tools}

During the preparation of this work, the authors used \texttt{Gemini} to generate the initial draft of this manuscript. After using this tool, the authors reviewed and edited the content as needed and take full responsibility for the content of the published article.

\bibliographystyle{elsarticle-num}
\bibliography{ref}

\end{document}